\documentclass[twocolumn]{aastex631}
\usepackage{multirow}
\usepackage{blindtext}
\usepackage{bm}
\renewcommand{\vec}[1]{\bm{#1}}
\usepackage{amsmath}

\accepted{April 17, 2023}

\shorttitle{G4CS II:  Characterization of 47\,\textit{Tuc} with Bayesian Statistics}
\shortauthors{Simunovic et al.}
\graphicspath{{./}{figs/}}

\begin{document}

\title{The GeMS/GSAOI Galactic Globular Cluster Survey (G4CS) II: Characterization of 47\,\textit{Tuc} with Bayesian Statistics.}

\correspondingauthor{Mirko Simunovic}
\email{mirko.simunovic@gmail.com}

\author[0000-0002-5652-6525]{Mirko Simunovic}
\affiliation{Subaru Telescope, National Astronomical Observatory of Japan, 650 North A'ohoku Place, Hilo, HI 96720, USA} 

\author[0000-0003-0350-7061]{Thomas H. Puzia}
\affiliation{Institute of Astrophysics, Pontificia Universidad Cat\'olica de Chile, Av. Vicu\~na Mackenna 4860, 782-0436 Macul, Santiago, Chile}

\author[0000-0002-5665-376X]{Bryan Miller}
\author[0000-0002-7272-9234]{Eleazar R. Carrasco}
\affiliation{Gemini Observatory/NSF’s National Optical-Infrared Astronomy Research Laboratory, Casilla 603, La Serena, Chile}

\author[0000-0002-4442-5700]{Aaron Dotter}
\affiliation{Department of Physics and Astronomy, Dartmouth College, Hanover, NH 03755 USA}

\author[0000-0001-5870-3735]{Santi Cassisi}
\affiliation{INAF – Osservatorio Astronomico di Abruzzo, Via M. Maggini, s/n, I-64100, Teramo, Italy}
\affiliation{INFN – Sezione di Pisa, Largo Pontecorvo 3, 56127 Pisa, Italy}

\author[0000-0002-9225-5822]{Stephanie Monty}
\affiliation{Research School of Astronomy \& Astrophysics, Australian National University, Canberra, ACT 2611, Australia}
\affiliation{ARC Centre of Excellence for Astrophysics in Three Dimensions (ASTRO-3D), Canberra 2611, Australia}

\author{Peter Stetson}
\affiliation{NRC, Herzberg Astronomy \& Astrophysics, 5071 West Saanich Road, Victoria, BC V9E 2E7, Canada}



\begin{abstract}

We present a photometric analysis of globular cluster 47 Tuc (NGC\,104), using near-IR imaging data from the GeMS/GSAOI Galactic Globular Cluster Survey (G4CS) which is in operation at Gemini-South telescope.~Our survey is designed to obtain AO-assisted deep imaging with near diffraction-limited spatial resolution of the central fields of Milky Way globular clusters.~The G4CS near-IR photometry was combined with an optical photometry catalog obtained from Hubble Space Telescope survey data to produce a high-quality color-magnitude diagram that reaches down to K$_s\approx$ 21 Vega mag.~We used the software suite BASE-9, which uses an adaptive Metropolis sampling algorithm to perform a Markov chain Monte Carlo (MCMC) Bayesian analysis, and obtained probability distributions and precise estimates for the age, distance and extinction cluster parameters.~Our best estimate for the age of 47 Tuc is 12.42$^{+0.05}_{-0.05}$ $\pm$ 0.08 Gyr, and our true distance modulus estimate is (m$-$M)$_0$=13.250$^{+0.003}_{-0.003}$ $\pm$ 0.028 mag, in tight agreement with previous studies using Gaia DR2 parallax and detached eclipsing binaries.

\end{abstract}



\section{Introduction} \label{sec:intro}

In the field of Galactic astronomy, Globular Clusters (GCs) remain one of the most important laboratories for testing models of stellar evolution and dynamics, as well as being authentic fossils that trace the Milky Way's formation and evolution processes.~In the last two decades, the study of Galactic GCs has gained enormous momentum from the technological advancements in ground observations as well as from space, more notoriously through the Hubble Space Telescope (HST).~In particular, the development at ground telescopes of adaptive optics (AO) opened the possibility to obtain diffraction-limited imaging in the dense central regions of GCs, and obtain accurate photometry from stars several magnitudes below the main-sequence turn-off (MSTO).~This kind of data is also available from HST imaging surveys of GCs, namely the HST ACS Survey of Galactic GCs \citep{sarajedini07} and the HST UV Globular Cluster Survey \citep[HUGS;][]{piotto15,nardiello18}.

Significant progress has been made in the task of measuring accurate GC ages, a field that historically suffered from many limitations including lack of suitable models, parameter degeneracy and insufficient high-quality data.~For example, the exquisite HST ACS photometry has been used to obtain ages based on precise isochrone model fits to the color-magnitude diagrams (CMD) of a large number of GCs \citep[e.g.][]{dotter10,paust10,wagner17}, as well as measuring relative ages using the MSTO region in the CMD \citep{marin09}, and distances based on the zero-age horizontal-branch (ZAHB) to better constrain the isochrone fits \citep{vandenberg13}.~The importance of accurate and precise age estimates is illustrated by the age-metallicity relation (AMR) of GCs, from which one can classify the GC population in, at least, two groups: (i) old clusters in the outer halo that span the full range of metallicities, and (ii) a sequence of GCs in the inner halo that become younger at higher metallicity \citep{rosenberg99,dotter10,vandenberg13,leaman13,wagner17}.~This empirical result gives ground to the idea that the inner halo formed rapidly, while the outer halo is likely populated by stars accreted from dwarf galaxies during merger events.~However, measuring the precise absolute age of globular clusters is necessarily subject to a set of assumptions in both the models  \citep[see e.g.][for a recent review of model physics]{joyce23} and the data \citep[see e.g.][for a discussion on ACS zero-point errors and isochrone fits]{brogaard17}, making it hard to find reproducible and systematic age measurements of Galactic GCs in the literature.

This is the second published work obtained from the \textit{GeMS/GSAOI Galactic Globular Cluster Survey} (G4CS).~The G4CS was designed mainly to obtain homogenous, deep near-IR J and K$_s$-band photometry of Milky Way GCs in order to obtain precise and accurate estimates of GC absolute ages.~This is possible by the use of the AO-assisted GSAOI \citep{mcgregor04,carrasco12} imager at Gemini-South telescope, which in tandem with the Gemini Multi-conjugate adaptive optics System \citep[GeMS;][]{rigaut14} is able to deliver near diffraction-limited imaging in the near-IR across the full 85$\times$85 arcsec$^2$ field-of-view of GSAOI.~In our first paper \citep[][hereafter Paper I]{monty18} we analyzed GSAOI data of NGC\,2298 and NGC\,3201 and showed the imaging capabilities of GeMS/GSAOI to reach about FWHM$\sim$0.08 arcsec of spatial resolution in the K$_s$-band and an astrometry precision of $\sim$0.2 mili-arcsec (mas).~Both levels of precision are comparable to those of HST and make therefore GeMS/GSAOI an optimal ground instrument for advancing the studies of resolved stellar populations in dense GCs.~For both target GCs in Paper I, we could constrain their absolute age to within less than 1 Gyr, making such estimates among the most statistically robust in the literature.

The more important advantages of conducting CMD analysis in the near-IR arise from the significantly lower line-of-sight absorption suffered at near-IR wavelengths (hence also minimizing the effect of differential reddening) as well as the lower main-sequence (MS) bending feature that appears in the CMD when including near-IR filters.~This feature is commonly referred to as the low MS ``knee" \citep[MSK;][]{calamida09,bono10,massari16} and is mainly caused by collisionally induced near-IR absorption of molecular hydrogen in the atmospheres of low-mass stars \citep{saumon94}.~Since the MSK is mostly age-independent for a fixed chemical composition, it is then possible to use the bending shape of the CMD to ``anchor" the isochrone models, hence also removing distance and reddening effects when combined with strong features like the MSTO.

On the other hand, the high-quality photometry and well-defined CMDs can provide isochrone fits that are only as good as our models, and as sensitive as the resolution of the discrete model grids that are commonly used.~Advanced statistical methods are therefore necessary to robustly quantify and characterize the best-fit cluster parameters.~In this paper, we follow previous similar works \citep[see e.g.][]{wagner16a,wagner16b,stenning16,wagner17,bossini19} and use Bayesian methods to explore simultaneously the model parameters (age, distance and extinction), while also incorporating into our model priors the information from previous studies.~This way, we can use Bayesian inference to obtain the joint posterior probability distribution of these parameters and understand their confidence intervals and covariances.

Due to its close distance and large stellar mass, 47 Tuc (NGC\,104) is a very well-studied Galactic GC and is usually used as a calibrator of metal-rich stellar population models, as well as a classical example of a metal-rich red horizontal-branch (HB).~We have used the G4CS near-IR imaging data of the Galactic GC 47 Tuc (NGC\,104) to obtain high spatial resolution and deep photometric data, allowing for a precise age estimate, and a new distance estimate that compares remarkably in agreement with the Gaia DR2 parallax distance.~

This paper is organized as follows.~In Section~\ref{sec:data_red} we present the data and data reduction, Section~\ref{sec:phot_analysis} details the photometry analysis and photometric catalogs, Section~\ref{sec:iso_fit} shows the Bayesian method of isochrone fitting and the results are given in Section~\ref{sec:results}. Finally, in Section~\ref{sec:discussion} we discuss the results in light of previous literature, and give a summary in Section~\ref{sec:conclusions}.

\section{Observations and Data Reduction}\label{sec:data_red}

\subsection{GSAOI Near-IR Data}
The imaging data was obtained during the night of December 10, 2017 with the Gemini-South 8-meter telescope on Cerro Pach\'on, Chile, under photometric and favorable seeing conditions.~These observations were part of an allocated G4CS proposal for the 2017B semester (Program ID: GS-2017B-Q-53, PI: T. H. Puzia) and were executed in Queue mode.~The instrument used was GeMS/GSAOI \citep{rigaut14,mcgregor04,carrasco12}.~The GeMS system uses five laser guide stars (LGS) and three natural guide stars (NGS) to illuminate different wave-front sensors (WFS) in order to correct the high-order atmospheric turbulence and the low-order tip-tilt distortion, respectively.~The AO-corrected near-IR light is then sent to be imaged by GSAOI.~The detector is a 2$\times$2 mosaic of Rockwell HAWAII-2RG which results in a 85" x 85" field-of-view, with a scale of 0.02" per pixel, and gaps between the arrays that project to $\sim$2.5" on the sky.

\begin{figure}[t!]
\includegraphics[width=0.5\textwidth]{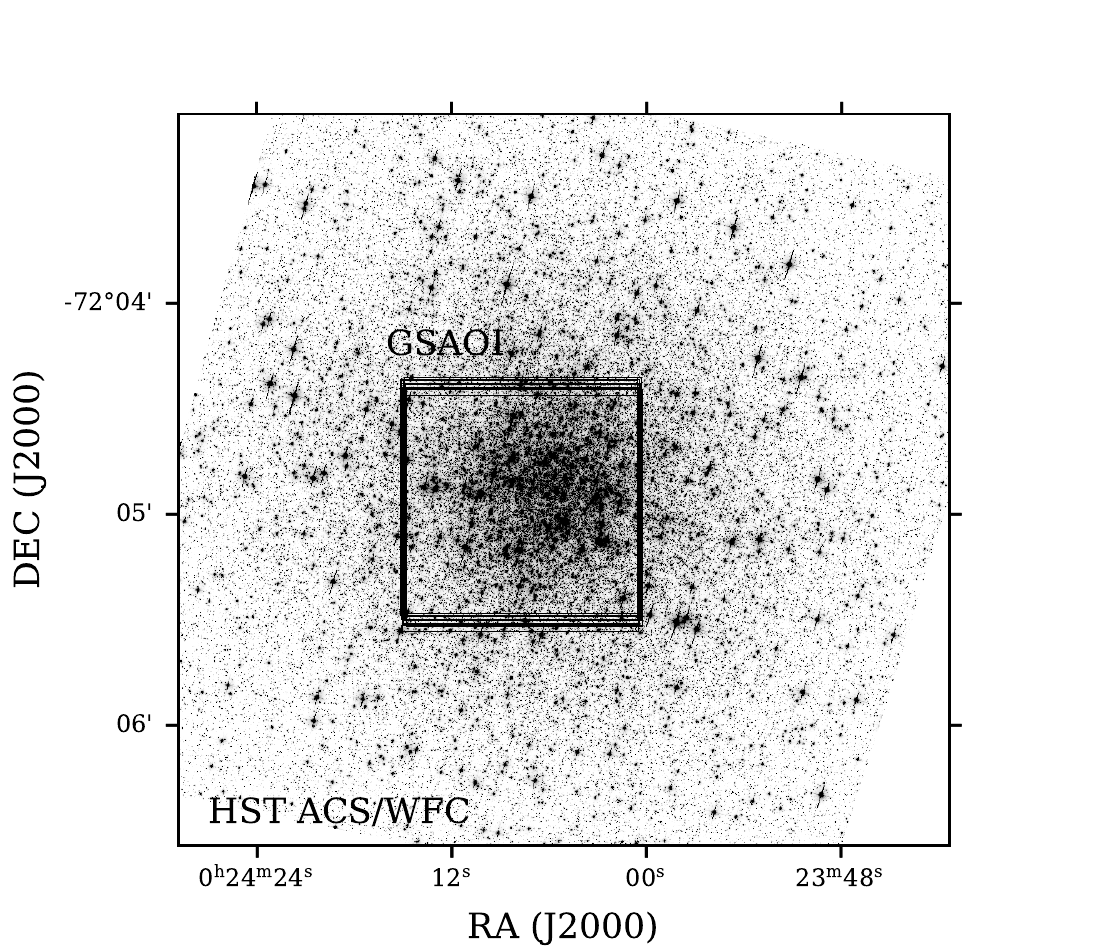}
\caption{HST ACS/WFC image of 47 Tuc from \citet{nardiello18}.~The field of view (85" x 85") and dither positions of the GSAOI data is shown at the center.  \label{fig:gsaoi_fov}}
\end{figure}

The target GC 47 Tuc was observed with the J and K$_s$ GSAOI filters by taking different coadd$\times$exposure combinations in order to increase the dynamic range of the measured sources (see Table~\ref{tab:observation}).~The sequence of exposures of each observing mode were arranged with a non-redundant dithering pattern that used offsets of a few arcseconds in each direction, which results in each pixel sampling a different part of the sky during each exposure.~This strategy allowed us to construct appropriate background frames to be used during the data reduction.~We show in Figure~\ref{fig:gsaoi_fov} an HST image of 47 Tuc from the HUGS survey realease \citep{nardiello18} and the region covered by our GSAOI images. 

\begin{table}
	\centering
	\caption{Observing Log for 47 Tuc during December 10, 2017. Program ID: GS-2017B-Q-53}
	\label{tab:observation}
	\begin{tabular}{cccc} 
		\hline
		\hline
		Filter & N$\times$coadd$\times$$t_{exp}$ & Natural Seeing & AO FWHM \\         
		\hline
		J & 5$\times$2$\times$6     & $\sim$0.5"$-$ 0.6"  & 110 $ -$180 mas \\
		   & 7$\times$2$\times$30 &  ''  &  140 $ -$ 170 mas  \\
		   & 6$\times$1$\times$60 & ''   & $>$200 mas   \\
	K$_s$ & 5$\times$2$\times$6 & ''  & $\sim$95 mas   \\
		   & 5$\times$2$\times$30 & ''   &  $\sim$95 mas  \\
		   & 5$\times$1$\times$60 &  ''   &   $\sim$90 mas \\		 		 		 
		\hline
	\end{tabular}
\end{table}

\begin{figure*}[t!]
\centering
\includegraphics[width=0.99\textwidth]{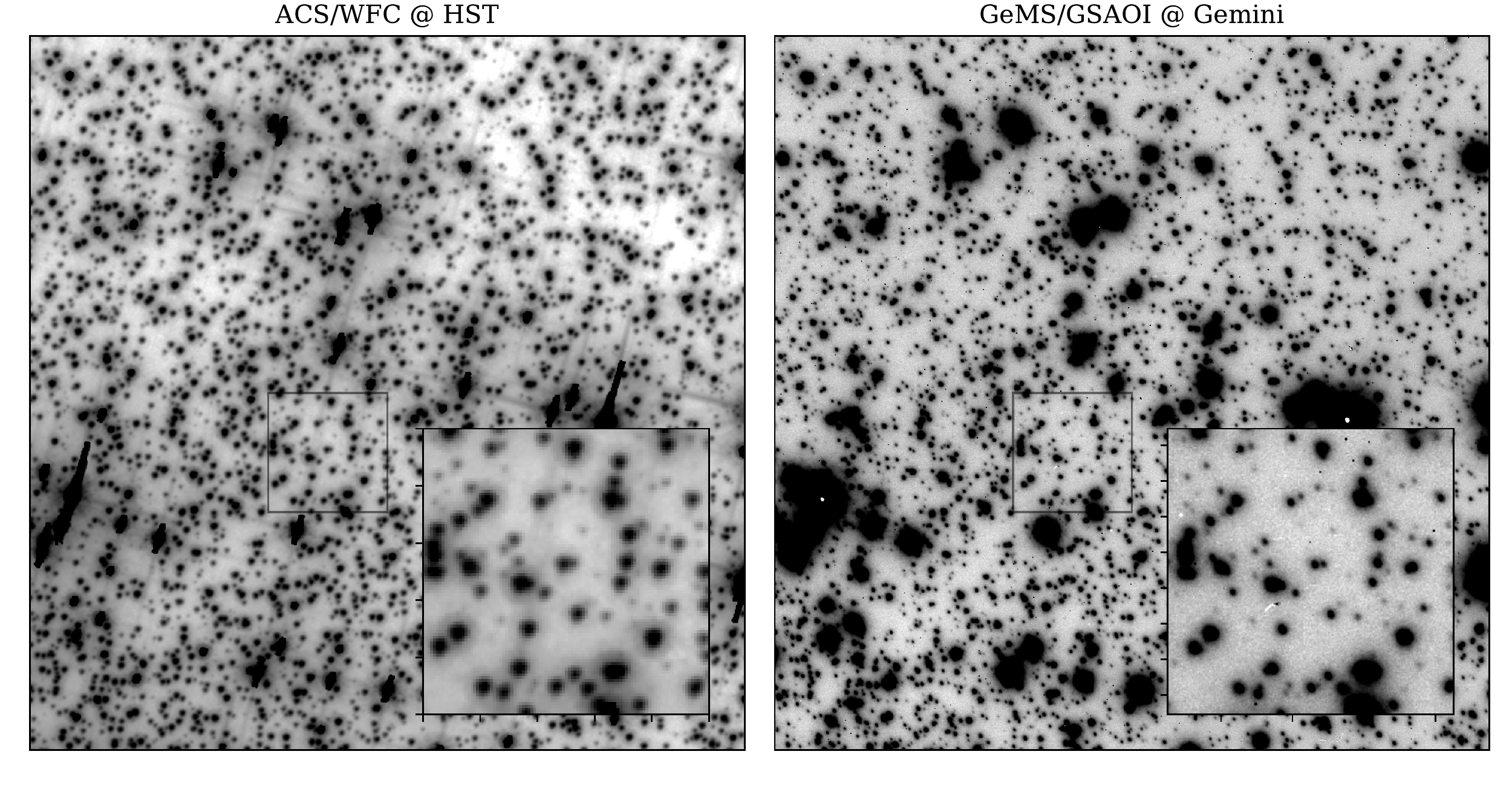}
\caption{Comparison of the spatial resolution of ACS/WFC camera of HST (left) with the GeMS/GSAOI camera at Gemini (right).~The images show identical 15$\times$15 arcsec$^2$ regions in the central field of 47 Tuc.~The inset panels show a close up of the regions marked by the square near the center.~The images are taken with the ACS F606W \citep{nardiello18} and GSAOI K$_s$ filters.      \label{fig:gsaoi_hst_comp}}
\end{figure*}

The raw images were accompanied by a set of standard Baseline Calibration frames for GSAOI, which correspond to dome flats taken with the lamps ON and OFF.~The OFF dome flats are needed for the K$_s$ images in order to correctly calibrate the thermal background in the flat field.~The full data reduction was managed with the THELI software package \citep{schirmer13,erben05} installed on a Linux virtual machine running Ubuntu 14.04.5. THELI\footnote{The software and instructions can be obtained at https://www.astro.uni-bonn.de/theli/gui/index.html} is a graphical user interface that can perform automated reduction of astronomical images for a large variety of instruments while taking full advantage of parallelization for optimal computing power.~The reduction follows the standard workflow for near-IR images in THELI.~First, all raw images are corrected for non-linearity using measured coefficients that are stored in the software.~The master flat is obtained by subtracting the median-combined OFF dome flat from the median-combined ON dome flat and then normalizing to unity.~The remaining steps were done for each set of filter and exposure time configuration (see Table~\ref{tab:observation}) separately.~All of the raw science images are flat fielded and calibrated for different gain levels.~The next step of background modeling proved to be the most delicate for our data set.~Due to the high stellar density in the field we combined the sequence of dithered images to create a static background model in each chip, with the two-pass modeling option within THELI.~The default parameters were used, and the $\mathtt{mask\, expansion\, factor}$ parameter was set between 4-5, in order to mask the faint extended halos of bright sources during the median combination.~The background models were subsequently smoothed using a kernel of size 350 and rescaled at the level of each image.~For each image, THELI updates its header with a precise WCS solution.~The WCS reference catalog comes from an HST astrometric-photometric catalog \citep[][see next sub-section]{nardiello18}, given its precise astrometry and distortion correction.~The resulting calibrated, flat-fielded, background-subtracted images were used separately in the photometric analysis, as we will describe in Section~\ref{sec:phot_analysis}.~We find that the imaging quality of the non-dithered individual calibrated exposures is well within the standards of the HST ACS camera, as shown in Figure~\ref{fig:gsaoi_hst_comp} for a K$_s$-band exposure of 2 coadd$\times$30 sec and a AO FWHM $\sim$90 mas.~Unfortunately, as Table~\ref{tab:observation} shows, the AO FWHM of the J-band images was significantly poorer, and thus we decided to reject them from our data, given that we cannot reach sufficient photometric depth nor signal-to-noise as required for our CMD analysis.~The following data reduction is therefore referring only to the K$_s$-band data.

\subsection{HST Optical Data}
In this paper we use a combined optical and near-IR CMD to obtain cluster parameters of 47 Tuc.~The optical HST photometry comes from the final data release\footnote{http://groups.dfa.unipd.it/ESPG/treasury.php} of the HUGS survey \citep{piotto15,nardiello18}, who in addition to the released near-UV HST data, also used the original ACS/WFC images from the HST ACS GC Survey \citep{sarajedini07} and performed improved photometry on the F606W and F814W data (Figure~\ref{fig:gsaoi_fov} shows the F606W data).~Some of the improvements include allowing temporal and spatial variation of the PSF models, neighbor flux subtraction, and a more careful treatment of saturated stars.~The adopted photometric catalog of 47 Tuc corresponds to the ``Method-1" photometry, as described in \citet{nardiello18}, which gives the best photometric quality over most of the magnitude range (see their Figure 3).~This HST catalog provides us with distortion-free ($x,y$) positions and Vega magnitudes in the near-UV and the optical (F606W and F814W). 

\section{Photometric Analysis} \label{sec:phot_analysis}

\begin{figure*}[t!]
\centering
\includegraphics[width=0.94\textwidth]{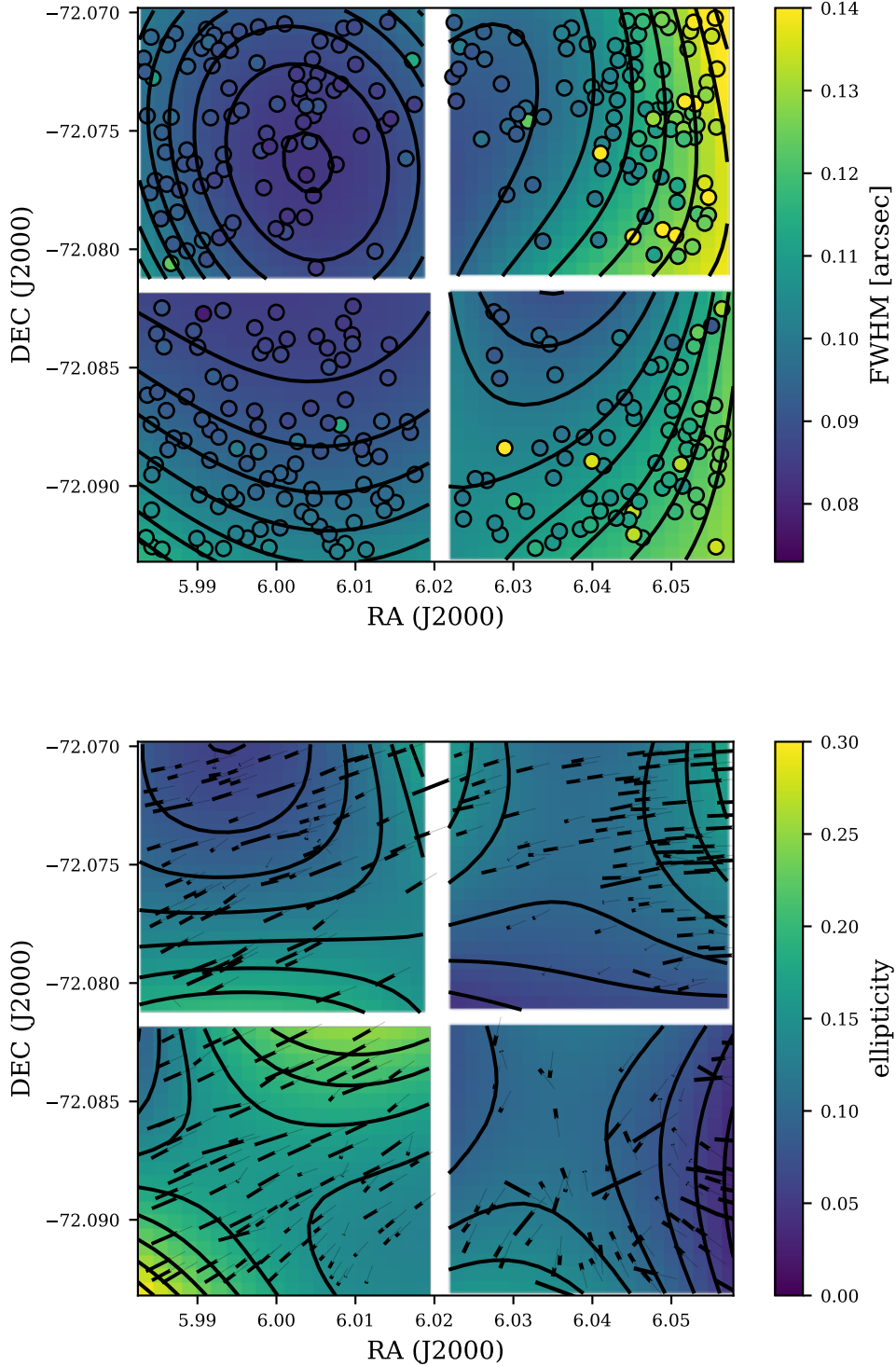}
\caption{Examples of the FWHM and ellipticity maps obtained in a 2 coadd $\times$ 30 sec K$_s$-band exposure. Top: Ra and Dec positions of the stars used to build the PSF model.~The measured FWHM of each star is shown in the color axis and a 2-d 3rd order polynomial is fit to the data in each chip, as shown by the contour plot. Bottom: Same as Top panel but for the case of ellipticity.~The position angle (PA) of each star is given by the angle of the bar, and its length is proportional to the ellipticity.  \label{fig:fwhm_ellip}}
\end{figure*}

\begin{figure*}[ht!]
\centering
\includegraphics[width=0.88\textwidth]{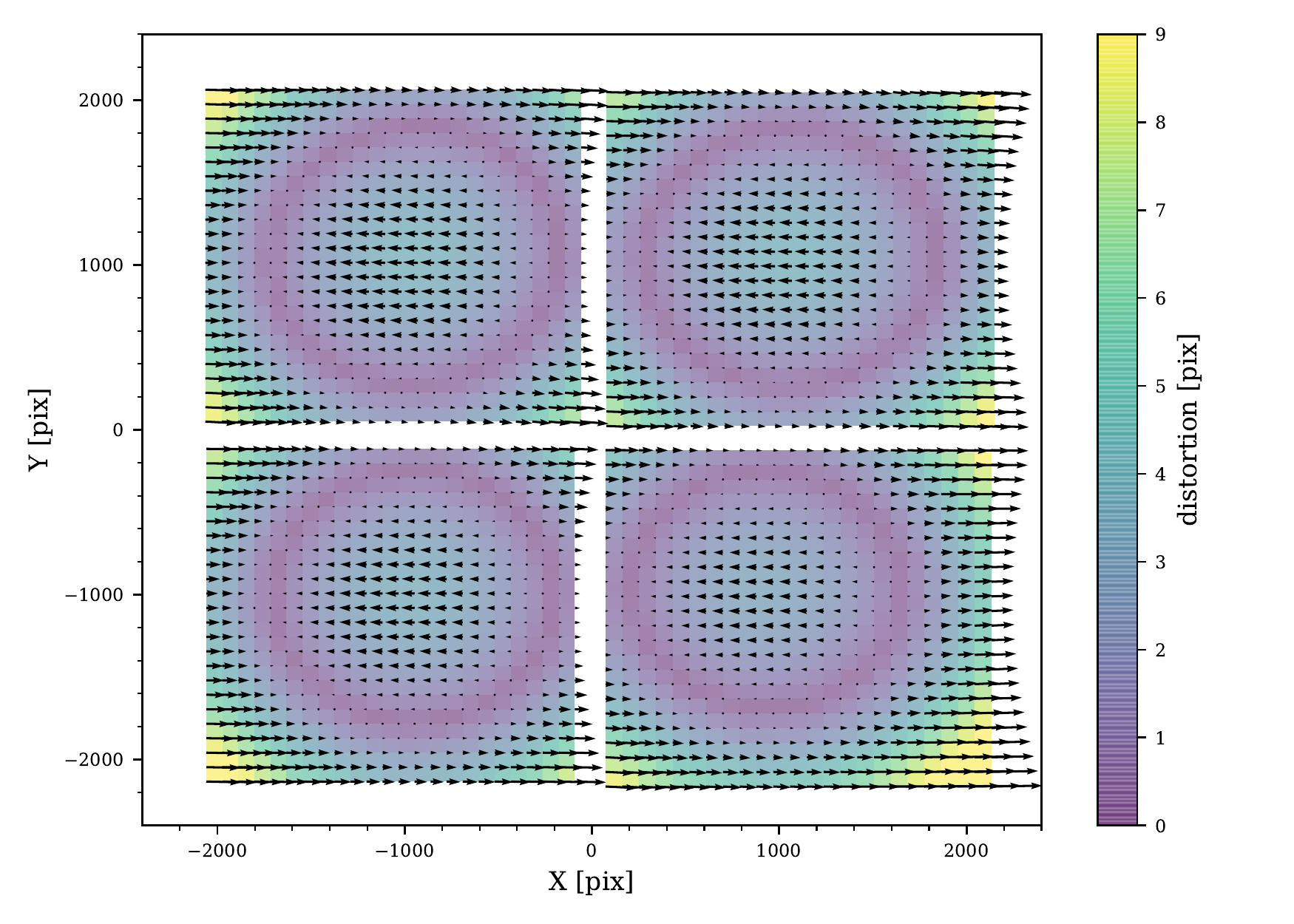}
\caption{Map of the geometric distortion measured in the GSAOI camera.~The distortion surface is modeled with a 3rd order polynomial fit to the residuals of the linear transformation.~The length of the vectors has been amplified by a factor of 20 for visualization.  \label{fig:geo_distortion}}
\end{figure*}

The performance of GeMS/GSAOI in dense stellar fields has been previously studied by different authors \citep{neichel14,massari16,dalessandro16,turri17,bernard18}.~It has been shown that the different distortions present in the camera optics of GeMS/GSAOI are hard to model due to their dependence on specific instrument configurations (e.g. offsets, relative position of the NGS and LGS constellation, etc.), and its PSF can change dramatically over the field-of-view.~In particular, the best values of FWHM are near the center of the LGS asterism, and the shape of the PSF (i.e. position angle and ellipticity) changes substantially across each detector field.~It is thus very important to be able to build a PSF model that is spatially variant with enough degrees of freedom.~For this we adopted the latest version of DAOPHOT/ALLSTAR \citep{stetson87,stetson94} which is capable of building a cubically variant PSF model, thanks to the large amount of available suitable stars in the field.~We performed photometry on each individual exposure, on a chip-by-chip basis, in order to obtain a model of the distortion and PSF-shape variation as thoroughly as possible.~The detected sources and their respective instrumental magnitudes were subsequently mapped into a non-distorted reference frame of coordinates and later cross-matched, and their magnitudes averaged.~The catalogs for each chip were then zero-point calibrated and combined to create the final photometric catalog for a single sequence of images taken with the same exposure time.~The procedure for obtaining accurate PSF photometry with DAOPHOT/ALLSTAR in crowded fields follows standard practices.~The stars used to model the PSF are monitored with the PSFMEASURE task in IRAF \citep{tody86} allowing us to iteratively build an improved PSF model across the chip (see Figure \ref{fig:fwhm_ellip} for an example of the PSF variability).~We use a Moffat function with $\beta$=3.5 for the analytic PSF model and a cubically variant model for the spatial dependence.

The selection of PSF stars is done with special care.~Their density and spatial distribution has a large effect on the precision of the final PSF model and therefore need to be chosen optimally.~Ideally we need PSF stars homogeneously distributed in the field, but this is unfortunately not always possible for the most densely populated regions of each chip, where the high stellar density prevents suitable and relatively isolated PSF stars from being found.~Usually we have $\sim$80-100 PSF stars in a single chip field (see Figure~\ref{fig:fwhm_ellip} for an example).~Yet, selecting a large sample of PSF stars comes at the cost of having to select a broader range of stellar magnitudes.~We thus ran a series of tests to determine what the optimum number of PSF stars is.~Note that this will be dependent on the instrument used and the stability of its PSF as well as the availability of bright isolated stars across the field.~For our 47 Tuc data, we find that typically after 60 stars the photometric precision reaches a plateau.~We still choose to work with all the available stars since it provides us with FWHM, ellipticity and position angle maps across the field.~The quantitative assessment of the photometric precision was performed with the method described in the next sub-section.

\subsection{Photometric quality using science-based metrics}\label{sec:phot_quality}
This is based on the work of  \citet{turri17}, who in their paper did a detailed analysis of the photometric performance of GeMS/GSAOI with DAOPHOT.~Most importantly, they studied the temporal and spatial variability of the PSF and discussed appropriate techniques to improve the obtained photometry with DAOPHOT.~As in their work, we assess the quality of our obtained photometry by measuring the width of the stellar loci in a CMD.~Unlike the cited work, we do not use the width of the MSTO but instead build a fiducial line in the region about $\sim$3 mag brighter and $\sim$2 mag fainter than the MSTO and measure the spread around it.~In particular, we use our HST photometric catalog of 47 Tuc and fit a gaussian distribution to the F606W$-$K$_s$ color-offset values (relative to the fiducial line), and the $\sigma$ of the gaussian distribution is used as the nominal quantitative measurement that is used to compare different parameter values in DAOPHOT and the overall photometry of the individual image.~This is a good approach in the sense that it uses science-based metrics, like the sharpness of the observed CMD, that are directly tied to our goal of obtaining the best isochrone fit to the observed CMD, while also using a single straight-forward observed quantity for an easy quality-control check of the sequence of steps in the photometric analysis. 

\subsection{Building the Master Chip Catalogs}\label{sec:master_catalogs}

\begin{figure*}[ht!]
\centering
\includegraphics[width=0.98\textwidth]{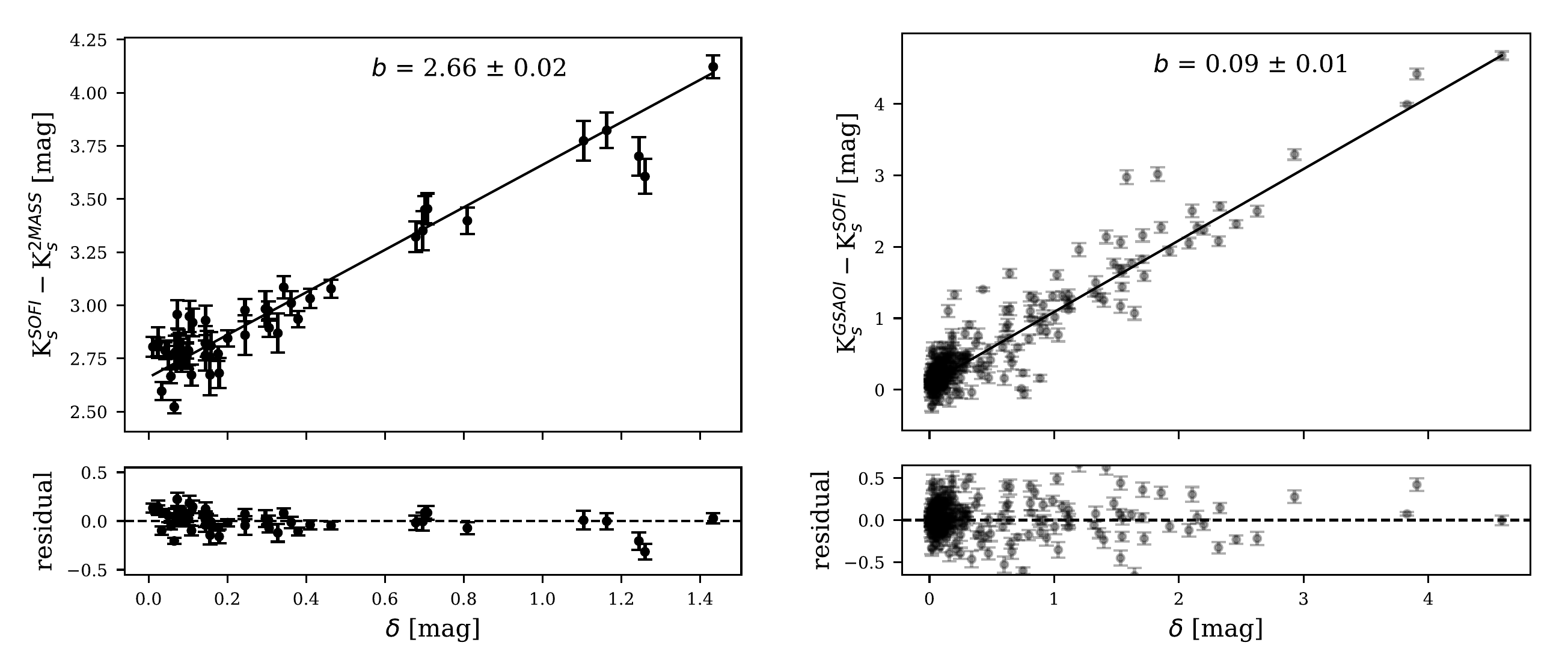}
\caption{Left: Zero-point calibration of the SOFI photometry into the 2MASS photometric system.~The x-axis shows the $\delta$ correction which arises from unresolved blending of stars in the 2MASS catalog.~The adopted zero-point for the particular exposure is given by the $b$ intercept parameter obtained from the linear fit (solid line) obtained for a fixed slope=1.~Right: Same as in the Left panel but showing the calibration of the GSAOI K$_s$ photometry into the SOFI (2MASS) photometric system.~The airmass extinction factor is already taken into account when doing this calibration. \label{fig:zp1}}
\end{figure*}

\begin{figure*}[ht!]
\centering
\includegraphics[width=0.98\textwidth]{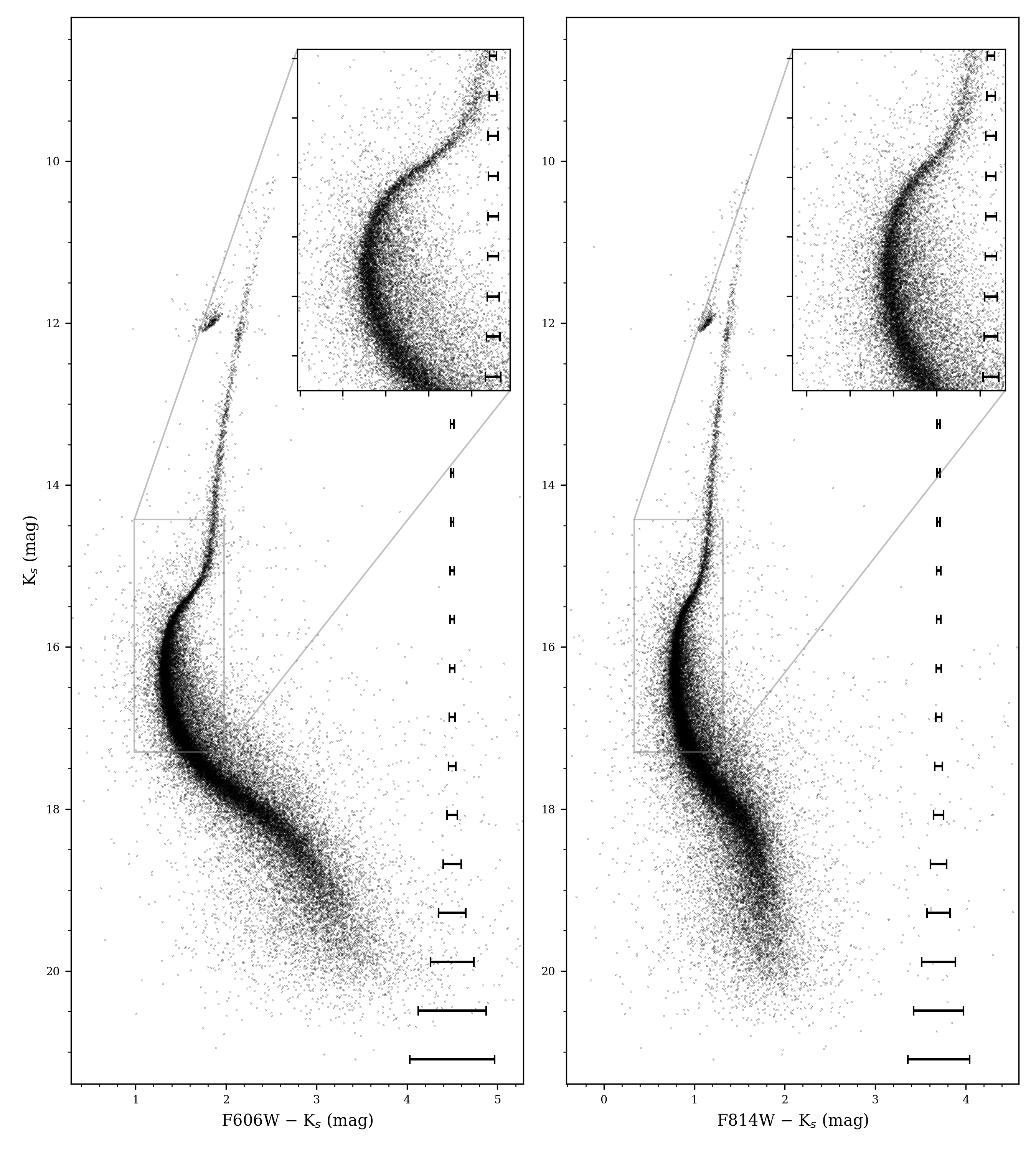}
\caption{CMDs of 47 Tuc for the combined HST optical filters and GSAOI K$_s$-filter photometry.~The inset panels show a zoomed plot of the MSTO and SGB regions in the CMD.~The horizontal error bars give the median photometric error in color at a given K$_s$ magnitude.    \label{fig:vk_cmd}}
\end{figure*}

The above procedure is used to obtain photometric catalogs of each chip for all images, per set of images at given exposure type.~The process of combining the catalogs into a master chip catalog is explained now.~We used the GEOMAP task in IRAF to map the $(x,y)$ coordinates of the stars in the GSAOI catalogs into a distortion-free coordinate frame.~For the reference catalog, again we used the ($x,y$) coordinates of the HST catalog.~The parameters used in GEOMAP were $\mathtt{fitgeometry=general}$, which includes a linear and a distortion term that are fitted separately.~The linear term has 6 coefficients and the distortion surface is modeled with a 3rd-order polynomial fit to the residuals, hence giving a total of 9 coefficients per axis.~This is the same approach we used in Paper I, where we showed that the obtained astrometric precision ranges between $\sim$0.2-1.0 mas for objects brighter than F814W$\approx$20 mag, in agreement with similar studies \citep{neichel14,massari16,dalessandro16}.~With this method, the geometric distortion across the GSAOI chips can be measured in great detail, as shown in Figure~\ref{fig:geo_distortion} for an example geometric transformation of one of our images.~The vectors show the residual distortion map obtained with GEOMAP and they have been amplified by a factor of 20 for better visualization.~The maximum distortion is $\sim$9 pixels around the outer edges of each chip.~While the map shown can only be applied to our specific data due to the variability of the distortions, it is qualitatively very similar to the distortion maps obtained recently by \citet{patti19} and \citet{bernard18} for GeMS/GSAOI.~The advantage of using the HST catalog for reference is that we can directly cross-match the sources and quickly produce a CMD combining the HST and GSAOI filters.~While the GSAOI magnitudes are not yet calibrated, the CMD is still useful to do a control-check of the photometric quality of the specific image and chip, as explained in Section~\ref{sec:phot_quality}.~The distortion-free coordinates are then used to cross-match objects accross the full sequence of GSAOI images and compute the median magnitude shift between different images, for each chip.~This shift is used to correct slight differences like variable seeing and atmospheric transparency between exposures.~We then do a more robust cross-match of objects using a 1 pixel (0.05 arcsec)\footnote{These are pixels in the distortion-free coordinate frame built from HST ACS/WFC coordinates, therefore 0.05 arcsec corresponds to a tolerance of 2.5 GSAOI pixels.} tolerance and build the final photometric catalog for each chip, obtained from the sequence of images.~All objects are stored in the master catalog and the number of detections will vary between 1 (e.g. stars in the edges of an image in the exterior of the dither pattern) and $N$, where $N$ is the number of images, while the stored magnitude is the error-weighed average for stars with more than a single detection.~The final photometric error is calculated as either the standard deviation of the magnitude between multiple detections, the difference between two detections, or the reported DAOPHOT photometric error for a single detection.

\subsection{Zero-point Calibration}\label{sec:zeropoint}

The obtained master catalogs are constructed for each individual chip, per set of images at given exposure type, and hence they need to be merged into a single master catalog.~This required precise zero-point calibration so that all catalogs have photometry in the same photometric system.~The GSAOI K$_s$-band magnitudes are calibrated to the 2MASS \citep{skrutskie06} photometric system, although an auxiliary photometric catalog should be used given that the 2MASS point-source catalog does not have enough stars in our field which are also not saturated in the GSAOI images.~We therefore used K$_s$-band imaging data obtained with the SOFI instrument, available at the ESO public archive (program  66.B-0247(A), PI: E. V. Held).~We used 3 exposures of $t_{exp}=5$ seconds observed at $\sim$0.8 arcsec seeing in a field near the center of 47 Tuc.~The raw images were reduced with THELI following a similar procedure as described in Section~\ref{sec:data_red} and the photometry was obtained with DAOPHOT/ALLSTAR on the stacked images.~The instrumental SOFI photometry is calibrated to the 2MASS photometric system with the equation
\begin{equation} \label{eq:zeropoint}
m_{2mass} = m_{inst} + Z +\kappa X
\end{equation}
where $Z$ is the zero-point, $\kappa$ is the atmospheric extinction coefficient and $X$ is the airmass.~The zero-point is obtained by comparing magnitudes in isolated stars in both catalogs.~However, the different spatial resolution between 2MASS (FWHM$\sim$3 arcsec) and the SOFI data (FWHM$\sim$0.8 arcsec) creates the problem that selected stars in the 2MASS catalog may include unresolved blends that are not taken into account by the 2MASS reported magnitudes.~This effect can be quantified by using the higher spatial resolution of the SOFI data and calculating the extra flux that is contributed from sources that lie within the PSF size of the 2MASS data.~Following the formulation described in \citet{turri17}, we can assume that the true magnitude of a reference star in the 2MASS catalog, m$_{t}$, can be expressed as
\begin{equation}\label{eq:delta}
 m_{t}  = m_{o}+\delta 
 \end{equation}
where m$_{o}$ is the \textit{observed} magnitude and $\delta$ is the correction additive constant (hence making the true magnitude fainter, as expected) that accounts for the unresolved stars that contribute to the observed flux.~We also need to define a radius $r_{nei}$ that sets the circlular area inside which we consider a neighbor star to be unresolved in the 2MASS images, which we define as $r_{nei}=1.35$ arcsec, or about $\sim$0.4FWHM for the 2MASS data.~This number is obtained by looking at the distribution of all nearest-neighbor distance values and selecting the highest possible before neighbor stars begin to be resolved.~Now, the reference star $i$ with a SOFI instrumental magnitude m$^i$ will have a $\delta$ correction that can be defined as
\begin{equation}
\delta=2.5\,log\left(1+\sum_j10^{(m^i-m^j)/2.5}\right)
\end{equation}
where m$^j$ is the instrumental magnitude of the $j$th "contaminant" star inside the circle.~Hence for all reference stars, any offset between the instrumental SOFI magnitude and the reference 2MASS magnitude will be defined by the zero-point plus the correction additive constant $\delta$.~This can be seen clearly in the left panel of Figure~\ref{fig:zp1} where we show a plot of the $\delta$ values in the x-axis and the offset between instrumental K$_s$ magnitudes and the reference 2MASS magnitudes in the y-axis.~Note that the relation is linear with a slope of 1, as expected from equation~\ref{eq:delta}.~Then, we do an error weighed linear regression setting the slope to 1 and obtain the zero-point as the y-axis intersect value, i.e. the magnitude difference for stars that would have zero contamination in the 2MASS data.~Note that the airmass extinction factor is already taken into account when doing this calibration.~Finally, we use the 2MASS (J$-$K$_s$) color-index to check for a color term and find no significant trend, as it was expected based on previous studies \citep{santos12}.
The same procedure was then applied to the GSAOI instrumental magnitudes, now using the calibrated SOFI catalog as reference and using $r_{nei}=0.24$ arcsec.~Similarly to the previous case, we show in the right panel of Figure~\ref{fig:zp1} the zero-point calibration which transform the GSAOI photometry into the SOFI (2MASS) photometric system.~Again, we ignore the color-term since we expect it to be insignificant based on our previous results from Paper I, as well as what has been reported by \citet{turri17} for the GSAOI filters.

\subsection{Final Catalog Merging}

The last step in building the final K$_s$-band photometric catalog is to combine the four master chip catalogs into a single master catalog, per type of exposure (i.e. short, medium and long).~After calibrating the K$_s$-band magnitudes as explained above we just append the four chip catalogs into a single table, and take the mean value of coordinate position, magnitude and error of any repeated sources (due to dithering).~This way we obtain three independent catalogs: from the set of short exposures, medium exposures and long exposures.~The final merging consists in cross-matching all stars available in the three catalogs and find the median magnitude offset relative to the short exposure catalog (which suffers less from non-linearity and saturation).~We apply this small (usually about $\sim$0.05 mag) offset and then obtain the mean values for coordinate positions, magnitudes and errors.~We observe that stars with K$_s$ $<$ 14 mag in the long exposure catalog show non-linearity when compared to the magnitudes of the short exposure catalog, and hence they were not used for the final merging.~The final K$_s$-band catalog now samples a much wider field compared to the individual chip catalogs, and thus an additional zero-point calibration is required because now we have many more stars available to use in the calibration.~The right panel of Figure~\ref{fig:zp1} is the actual calibration of the final merged catalog, which explains why the obtained zero-point is only 0.09 mag, given that the earlier pre-merging catalogs were already calibrated as well. 

The final K$_s$-band catalog was joined with the 47 Tuc HST catalog from \cite{nardiello18} by cross-matching the mapped distortion-free ($x,y$) coordinates and using a pixel tolerance of 0.5 pixels.~We show in Figure~\ref{fig:vk_cmd} the (F606W$-$K$_s$) and (F814W$-$K$_s$) CMDs for stars with $N_{det}>2$, where $N_{det}$ is the number of individual exposure detections from the GSAOI imaging data.~Because of the broader spectral baseline given by the F606W$-$K$_s$ color, and the overall better photometry offered by the F606W magnitudes, we will use F606W$-$K$_s$ for our definitive optical-nIR CMD.~While the possibility of using additional HST magnitudes for the isochrone fitting is interesting (e.g. in the near-UV), we choose to adopt only these filters because stellar models are far better constrained in optical wavelengths and also incongruences in the HST zero-points can effectively introduce uncalibrated systematics in our results \citep[see][for a discussion on zero-point incongruences in ACS magnitudes and their effect on the isochrone fits]{brogaard17}.

\section{Bayesian Isochrone Fitting}\label{sec:iso_fit}
For the isochrone fitting method, we use the \textit{Bayesian Analysis for Stellar Evolution with Nine Parameters} software (BASE-9)\footnote{code and manual available at https://base-9.readthedocs.io/en/latest/}.~This is a Bayesian software suite that recovers star cluster and stellar parameters from photometry and that was originally developed by \citet{vonhippel06}.~The power of BASE-9 relies on its use of an adaptive Metropolis (AM) Markov chain Monte Carlo (MCMC) sampling technique developed by \citet{stenning16}. The AM sampling is designed to avoid fine-tunning of the mulit-dimensional proposal distribution (due to correlation on the cluster parameters) and it therefore optimizes the proposal distribution by using the entire chain history to adapt in each iteration.~The reader is referred to \citet{stenning16} for a detailed description of the AM MCMC algorithm and its mathematical foundations.

The general description of the Bayesian inference in BASE-9 is to adopt a cluster model $G(\vec{M},\vec{\Theta})$, in which the vector 
\begin{equation}
\vec{\Theta}=(\theta_{log(age)},\theta_{[Fe/H]},\theta_{Y},\theta_{(m-M)_V},\theta_{A_V})
\end{equation}
defines the cluster global parameters and $\vec{M}=(M_1..M_N)$ is the vector that contains the zero age stellar masses for $N$ stars.~Then the vector of predicted magnitudes of the cluster model is $\vec{\mu}=G(\vec{M},\vec{\Theta})$.~For $N$ stars and $n$ filters, we have a set of observed magnitudes $\vec{X}$ with elements $x_{ij}$ for the $i$-th star and the $j$-th filter, and known gaussian (independent) errors defined by $\vec{\Sigma}$ with elements $\sigma_{ij}$.~We can then write the likelihood as
\begin{multline}
p(\vec{X}|G,\vec{\Theta},\vec{M},\vec{\Sigma},\vec{Z}) =\\ \prod_{i=1}^{N}[Z_i\times p(\vec{X}_i|G,\vec{\Theta},M_i,\vec{\Sigma_i},Z_i=1) \\ + (1-Z_i)\times p(\vec{X}_i|Z_i=0) ]
\end{multline}
where the $i$-th star can have a value $Z_i=1$ or $Z_i=0$ for the case of cluster member or field star, respectively.~In BASE-9, the magnitude of field stars is modeled with a constant uniform probability across the magnitude range of the input data, so that $p(\vec{X}_i|Z_i=0)=c$ and zero elsewhere.~Assuming then that the observed data is a normally distributed random variable, i.e. $x_{ij} \sim N(\mu_{ij},\sigma_{ij}^2)$, we can thus give the explicit likelihood for the $i$-th cluster member star
\begin{multline}
p(\vec{X}_i|G,\vec{\Theta},M_i,\vec{\Sigma_i},Z_i=1)=\\  \frac{1}{\sqrt{(2\pi)^n|\vec{\Sigma_i}|}}exp\left(-\frac{1}{2}(\vec{X_i}-\vec{\mu_i})^T\vec{\Sigma_i^{-1}}(\vec{X_i}-\vec{\mu_i})   \right)
\end{multline}
so that its magnitude vector (for $n$ filters) is modeled with a multivariate Gaussian probability distribution.~We note that the formal description of the likelihood in BASE-9 allows for binaries and thus contains the mass ratio parameter $R_i$.~In this study we use the $\mathtt{noBinaries}$ flag and thus all stars are treated as singles.~We expect 47 Tuc to have low binary fraction \citep[$f_b$=0.018;][]{milone12} in our field  and so neglecting the binaries does not have a significant effect on the results (see Section~\ref{sec:binary_test} for details). 

The MCMC sampling then uses an iterative approach to explore the joint posterior distribution $p(G,\vec{\Theta}|\vec{X},\vec{\Sigma})$, in which the ``nuisance" parameters ($\vec{M},\vec{Z}$) have been marginalized, thus producing a sample, or chain, of correlated draws from the joint posterior distribution.~The advantage of the AM iteration used in BASE-9 is that it adapts to the observed correlations in previous iterations to ensure that the sampling is more efficient and needless of fine-tuning by the user.~More details about BASE-9 and the statistics background can be found in the referred literature \citep{vonhippel06,degennaro09,vandyk09}.

Following recent similar work by \citet{wagner17} and \citet{bossini19}, we use BASE-9 to fit a single simple stellar population model and simultaneously fit the age, observed distance modulus (m$-$M)$_V$, and line-of-sight absorption $A_V$ of 47 Tuc and obtain their posterior probability distributions.~The metallicity and helium fraction were kept as fixed parameters by relying on the broad literature of high-resolution spectral line abundances for the case of metallicity \citep{alvesbrito05,mcwilliam08,koch08,carretta09c,carretta10,gratton13,thygesen14,cordero14,colucci17}, as well as the best literature available for the helium abundance \citep{briley97,anderson09,milone12,dicriscienzo10,nataf11,salaris16,denissenkov17}.~The decision to keep these fixed is based on the fact that the model grid used in this study does not have the required step sensitivity, particularly on helium, and the fine sampling required by BASE-9 could fall at risk of discontinuities. 

The BASE-9 module that we use is $\mathtt{singlePopMcmc}$ and some initial parameters need to be set by the user.~The number of iterations for the burn-in stage is set to the default value (1,000) and the number of MCMC iterations (after burn-in) is set to 10,000.~For the models, we use Dartmouth Stellar Evolution Database (DSED) isochrones \citep{dotter08} which are readily available in the BASE-9 software.~Unfortunately, no additional models from the BASE-9 library could be used for comparison because only the DSED isochrone files provide magnitudes in the HST ACS/WFC as well as 2MASS K$_s$ filters.\footnote{~Although other models have been designed to include these filters (e.g. BaSTI), the required files are not available in BASE-9 as off-the-shelf models. }

~Next, starting values and priors have to be defined for all the cluster parameters.~The priors are defined as gaussians centered on the starting value of the chain and having a specific gaussian $\sigma$ as described below:

\textit{Metallicity}: For the subsequent analysis and discussion we choose to adopt the global metallicity [M/H], or log($Z/Z_{\odot}$), since $Z$ is the actual physical parameter that stellar models use as input.~This also serves us to bypass the fact that the DSED models in BASE-9 are available only for [$\alpha$/Fe$]=0.0$, and hence the iron abundance [Fe/H] parameter in BASE-9 cannot be taken at face value because the metal mixture in 47 Tuc  is not solar scaled (see Table~\ref{tab:metal_literature} and Section~\ref{sec:metal_content} for details).~For this reason the [Fe/H] output values in BASE-9 are converted following \citet{salaris93}:
\begin{equation}\label{eq:metal}
[M/H]=[Fe/H]+log\left(0.638\times10^{[\alpha/Fe]}+0.362\right)
\end{equation}
 which for the case when [$\alpha$/Fe$]=0.0$, we simply get that [M/H] = [Fe/H].
 
 The metallicity prior is then a fixed [M/H] value with zero dispersion.~We adopt the value [M/H]$=-0.5$ based on a collection of high-resolution spectral abundance studies of the metal content of 47 Tuc (see Table\,\ref{tab:metal_literature}).~More on this in Section~\ref{sec:metal_content}.
 
\textit{Helium Mass Fraction}: A review of the relevant literature shows consensus on the existence of a helium spread in 47 Tuc up to $\Delta Y \sim0.03$ \citep{briley97,anderson09,milone12,dicriscienzo10,nataf11,salaris16,denissenkov17}, which is centrally segregated and associated with second generation stars.~A recent study using state-of-the-art HB models found that $Y$=0.287 for the most central regions of the cluster \citep{denissenkov17}.~We then use $Y$=0.28 as our fixed prior, but the reader is warned that an exact value is to be taken with caution, since the helium enrichment is a function of the cluster radius \citep{gratton13,li14}.~A more detailed discussion is given in Section~\ref{sec:helium_content}. 

\textit{Distance Modulus}: We use $(m-M)_V=13.37$ mag from \citet{harris10} as starting point and $\sigma(m-M)_V=0.1$ for the prior distribution, to reflect the broad range of distance estimates in the literature (more on this in Section~\ref{sec:distance_and_age}).

\textit{Absorption}: The previous literature seems generally in agreement for reddening values between E(B$-$V) = 0.03 (dust map of \citet{schlegel98}) and 0.04 \citep{harris10}.~For a standard extinction law ($R_V=3.1$) this corresponds to $A_V\sim0.09-0.12$ and so for simplicity we just adopt $A_V=0.1$ for the starting value and a dispersion $\sigma(A_V)=0.05$. 

For the age, we use a uniform prior in order not to bias the posterior in any way.~The starting value is arbitrarily set to log(age) = 10.10, although we found this to have no effect on the results.~The summary of the priors and starting values are shown in Table~\ref{tab:priors}.

Finally, to keep computation times reasonable ($\sim$10 hrs) we select a set of 10,000 stars from the F606W and K$_s$ magnitude catalog, while keeping equal number of stars above and below the MSTO.~This type of selection is needed to ensure that enough stars are available at the SGB and MSTO regions, where the diagnostic power for age measurement is significant.~This will correspondingly change the obtained luminosity function (LF) of the sample, but our fitting procedure is mostly dependent on the shape of the CMD, and therefore not affected directly by the LF provided that enough stars are available to accurately trace the observed CMD shape.~As a final filter we remove all stars that clearly belong to the HB and binary sequence in the CMD.~The CMD of our selected sample used in BASE-9 is shown as the black points in Figure~\ref{fig:vk_isochrone}.~From the Figure we can also distinguish that the lower MS region, and in particular the expected MSK feature is not well defined in our data for it to be significant during the statistical modeling.~In fact, as noted by \citet{saracino18} in a recent study, the optical-nearIR CMD is not capable of accurately revealing the MSK and this can be only achieved using pure near-IR CMDs.~Moreover, \citet{saracino18} also found $\sim$0.2 dex variability in the MSK magnitude among different stellar models that cannot be calibrated from the systematic errors.~We therefore emphasize that in this study we do not particularly rely on the precise structure of the MSK, and the model uncertainties around the lower MS become unimportant given the poorer photometric quality of the lowest MS stars which effectively widens the parameter posteriors in favor of the CMD regions better constrained by our data.

\begin{table}
	\centering
	\caption{List of input parameters used in BASE-9 for the MCMC sampling.}
	\label{tab:priors}
	\begin{tabular}{rrrr} 
		\hline
		\hline
		Parameter & \multicolumn{2}{c}{Prior}  & Starting value \\
		                   &   center   & dispersion &     \\
		 
		\hline
		  [M/H]   &  $-0.5 $ & 0 (fixed) & $-0.5$ \\
		  $Y$      &   $0.28$  &  0 (fixed) & $0.28$ \\
	        $(m-M)_V$  &  13.37  & 0.1 & 13.37  \\
	        $A_V$     &  0.1    &  0.05    & 0.1 \\   
	        log(age)       & \multicolumn{2}{c}{uniform}    &  10.10     \\
		\hline
	\end{tabular}
\end{table}

\begin{figure*}[ht!]
\centering
\includegraphics[width=0.9\textwidth]{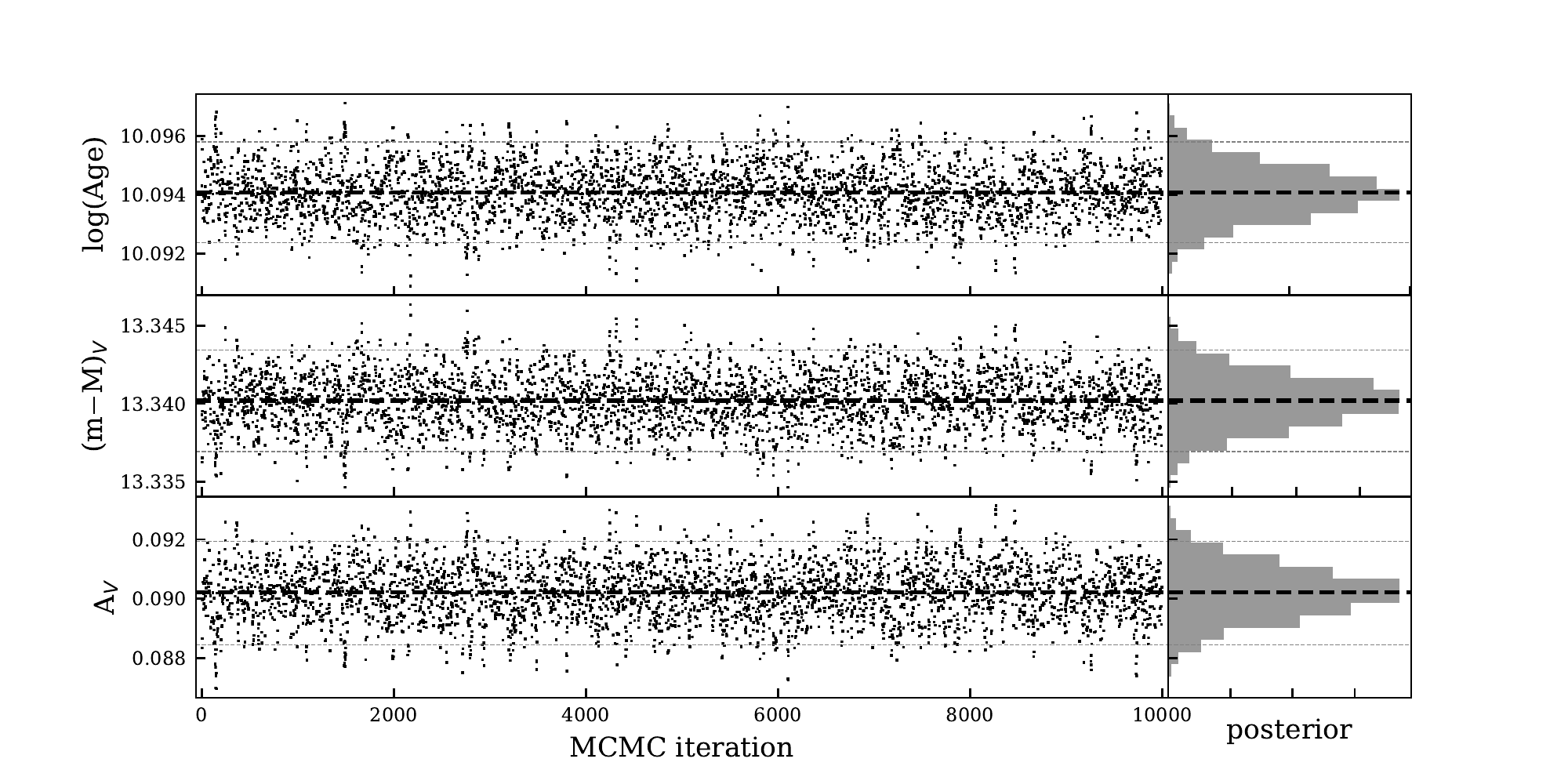}
\caption{MCMC chain of all the model parameters as labeled in the y-axis.~The black horizontal dashed lines give the median value of the posterior distribution and the light dashed lines give the 95\% Bayesian credible intervals.   \label{fig:post_dist}}
\end{figure*}

\begin{figure*}[ht!]
\centering
\includegraphics[width=0.62\textwidth]{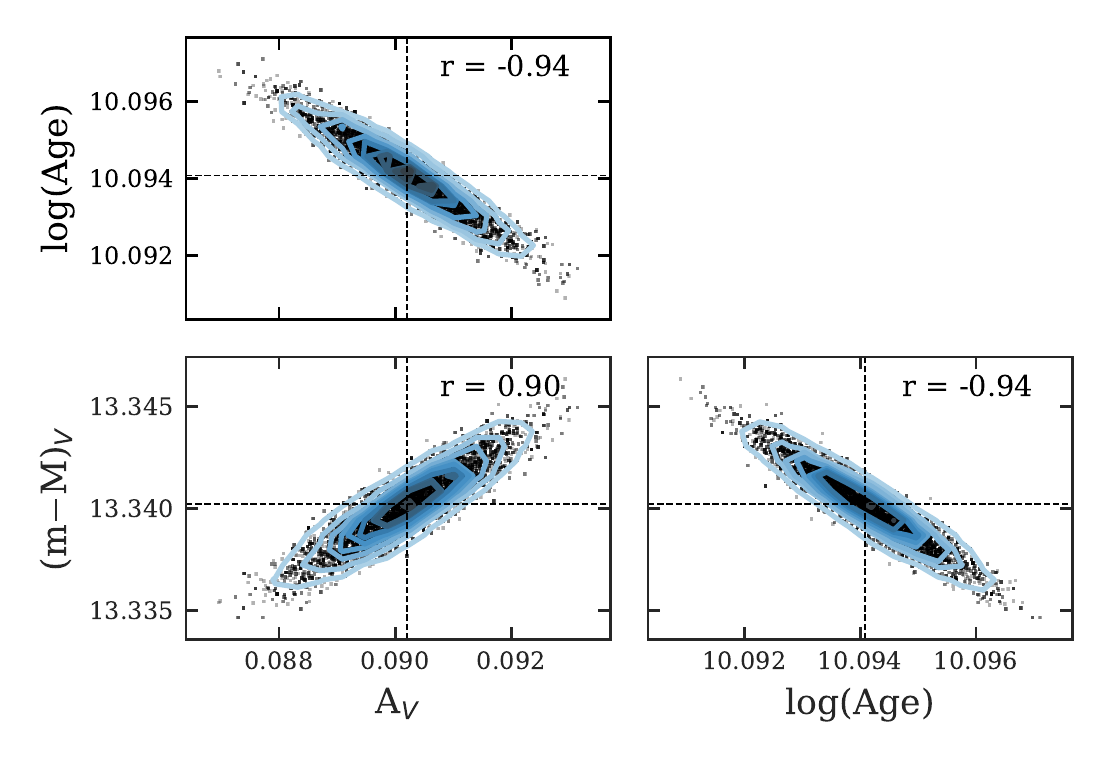}
\caption{Two-parameter covariance plots of the posterior draws from the MCMC sampling for all possible combinations.~The dashed black lines show the median of the posterior distribution.~In each scatter plot, the contours show a 2-d kernel density estimator that is fitted to the data points.~The Pearson correlation coefficient is given in the top right. \label{fig:joint_post}}
\end{figure*}

\section{Results}\label{sec:results}

\begin{table}
	\centering
	\caption{Obtained parameters for 47 Tuc from MCMC sampling.}
	\label{tab:base9_results}
	\begin{tabular}{ccc} 
		\hline
		\hline
		\multicolumn{3}{c}{3-parameter model MCMC posteriors*}\\
		\hline
		 Age [Gyr] & $(m-M)_V$ [mag]& $A_V$ [mag] \\
		\hline
                 12.42$^{+0.05}_{-0.05}$ & 13.340$^{+0.003}_{-0.003}$ & 0.090$^{+0.002}_{-0.002}$ \\ \\                                             
                 
		\hline
		\multicolumn{3}{l}{\small *The central estimate value is taken from the median}\\
		\multicolumn{3}{l}{\small of the posterior distribution. Errors correspond to the}\\
		\multicolumn{3}{l}{\small 95\% Bayesian credible intervals.}
		\end{tabular}
\end{table}

\subsection{MCMC Sampling}\label{sec:mcmcresults}

\begin{figure}[ht!]
\centering
\includegraphics[width=0.48\textwidth]{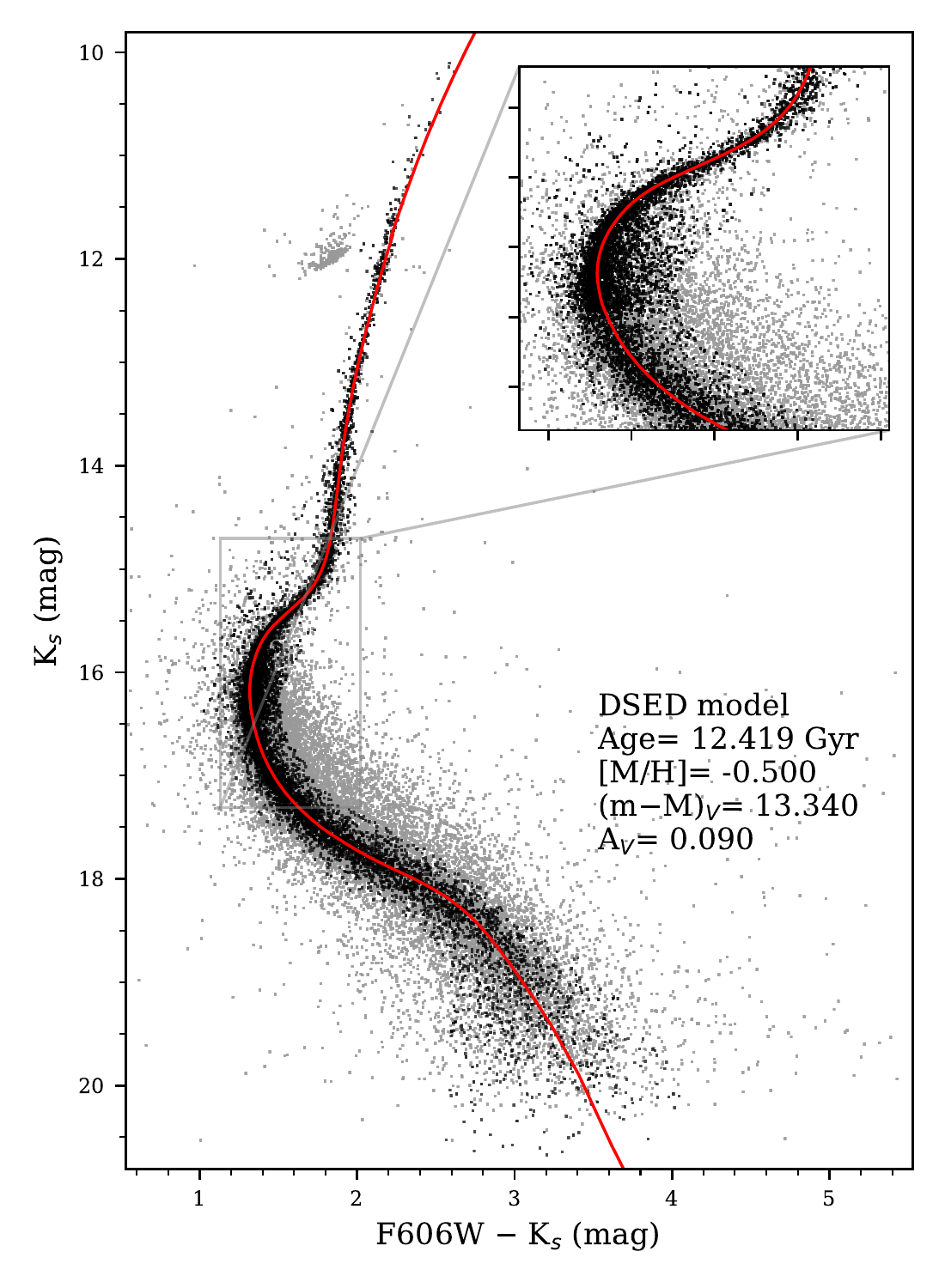}
\caption{CMD of 47 Tuc showing the stars used in the analysis with BASE-9 (black points).~The DSED isochrone created with the adopted cluster parameters from MCMC sampling is shown for comparison.~The isochrone can correctly match the entire CMD, including the sharp SGB and MSTO regions as seen in the inset panel.    \label{fig:vk_isochrone}}
\end{figure}

The values obtained from the 10,000 MCMC iterations are shown in Figure~\ref{fig:post_dist}.~For each fitted parameter, we show its posterior distribution along with the median values.~The 2d correlations are also shown in Figure~\ref{fig:joint_post} for all different combination of GC parameters.~We adopt the median of the posterior distribution as the nominal value for each cluster parameter and the quoted errors correspond to the 95\% Bayesian credible interval.~Note that these errors represent only the random statistical uncertainties.~We show the results in Table~\ref{tab:base9_results}.~The DSED isochrone that corresponds to the adopted cluster parameters is also shown in Figure~\ref{fig:vk_isochrone}.~Note that the isochrone can very accurately match the entire CMD structure, including the sharp SGB and MSTO regions as seen in the inset panel.~We find that both the estimated age and inferred heliocentric distance are near the average of a collection of literature values (see Figure~\ref{fig:age_dist_lit} for further details).~We discuss our findings more in detail and compare with previous literature in Section~\ref{sec:discussion}.~We want to highlight the interesting correlations that are visible from the joint posterior distributions in Figure~\ref{fig:joint_post}.~The Pearson correlation coefficient $r$ is shown on the top right of each panel.~Note that all parameters are strongly correlated.~The sign of the correlation is as expected, since older isochrones require a fainter MSTO, and thus a closer distance and/or smaller absorption is needed to match the data.

\subsection{Effect of Binaries}\label{sec:binary_test}
\begin{figure}[t!]
\includegraphics[width=0.47\textwidth]{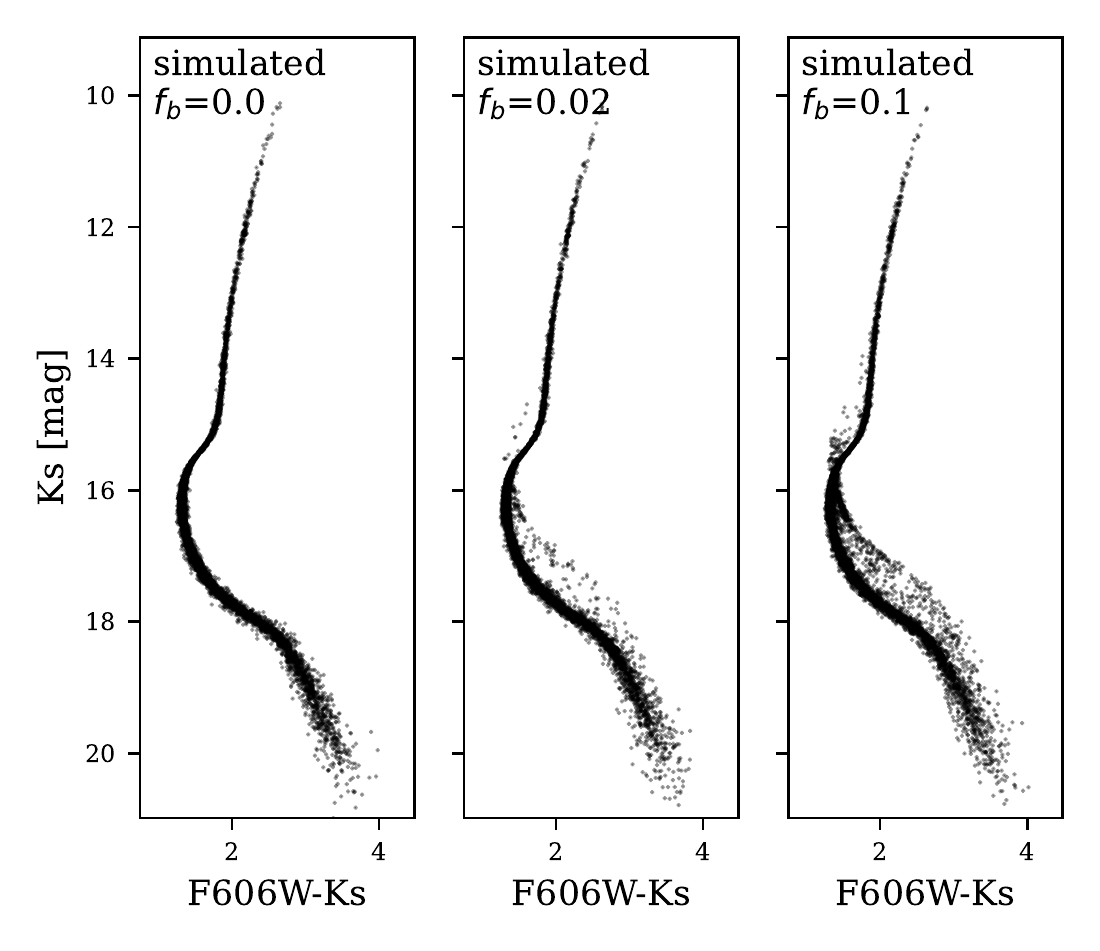}
\caption{CMDs of mock photometric catalogs simulated with BASE-9 for a set of cluster parameters equal to the adopted values for 47 Tuc.~Three cases were simulated for different binary fractions, as shown in the top of each panel. \label{fig:bin_test_cmd}}
\end{figure}

\begin{figure}[t!]
\centering
\includegraphics[width=0.45\textwidth]{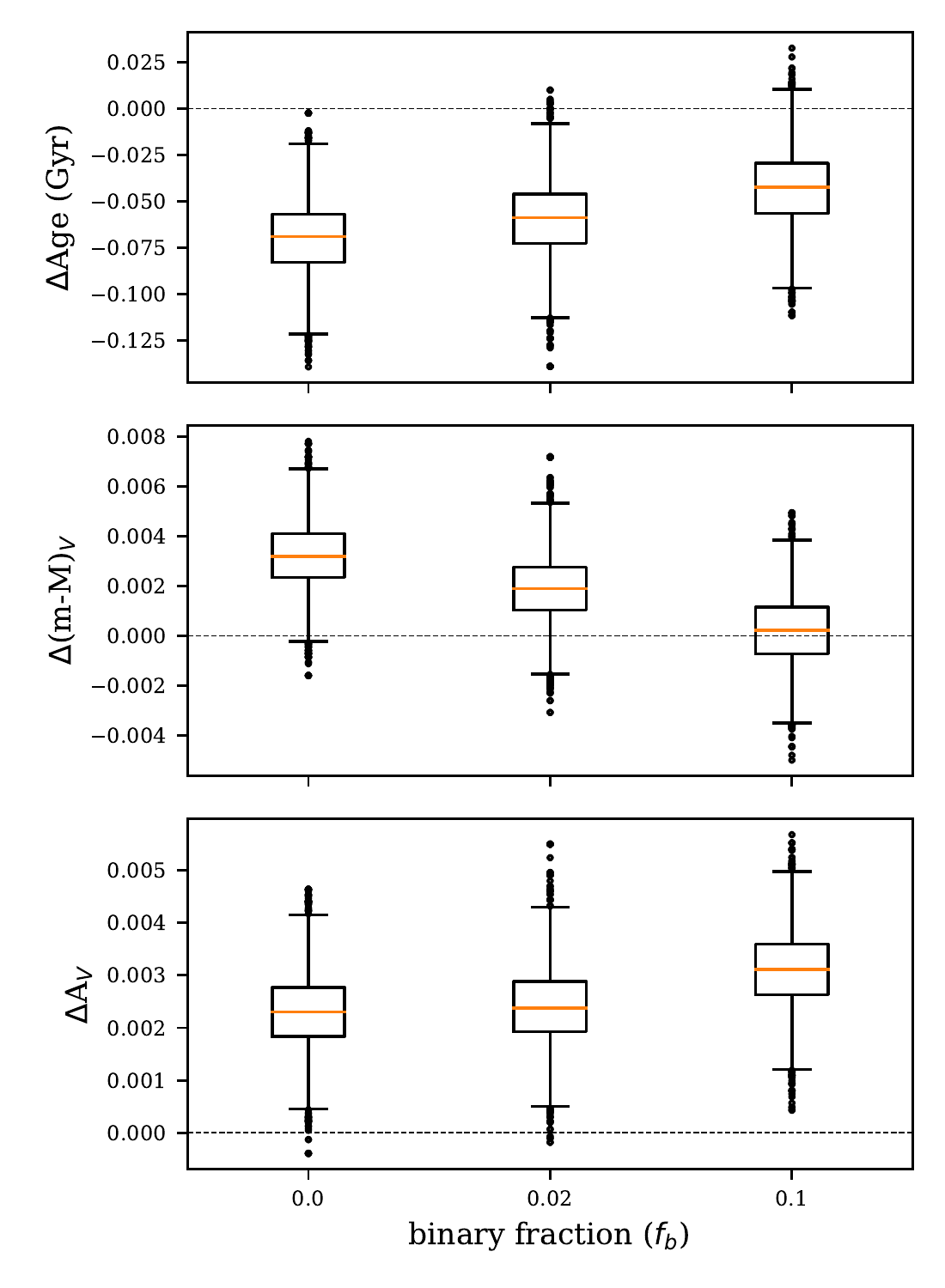}
\caption{Box plots of the derived posterior distributions for simulated catalogs assuming different binary fractions.~The y-axis gives the offset defined as $\Delta\theta = \theta-\theta_{true}$.~The horizontal dashed line is then perfect agreement between the true parameter and the posterior.~See Section~\ref{sec:binary_test} for details. \label{fig:bin_test_posterior}}
\end{figure}

We have ignored the binary population and hence $\mathtt{singlePopMcmc}$ was executed with the binary parameter turned off.~It is therefore necessary to measure the effect that the exclusion of binaries in the models can have on the MCMC results from previous sections.~For this test, we take DSED models and use the $\mathtt{simCluster}$ tool of BASE-9 to simulate a stellar population with N=20,000 stars and the cluster parameters obtained in section \ref{sec:mcmcresults}.~Three cluster models with a respective photometric catalog (including F606W and K$_s$ magnitudes) are created, each with binary fractions $f_b$= 0.0, 0.02 and 0.1, respectively.~Note that we expect 47 Tuc to have a relatively small binary fraction in the central regions: $f_b$= 0.018 as measured from HST photometric data \citep{milone12b}.~For each model, we further reduce the sample to 10k stars by copying the K$_s$-filter luminosity function found in the set of stars that was used in section \ref{sec:mcmcresults}, which we will refer to as the real data sample.~In addition, for each of the 10k stars we define a 0.2 mag region centered on its F606W magnitude, and then select a unique random star (which has not been selected before) from the real data sample inside that region.~The photometric error in F606W of the real star is then assigned to the star in the simulated catalog, and subsequently the noise is added to the simulated magnitude by drawing from a normal distribution with its standard deviation equal to the assigned photometric error.~This process is then repeated for the K$_s$-filter magnitudes, and thus we end up with photometric catalogs suitable to be fed into BASE-9 for analysis.~The CMDs of the simulated catalogs are shown in Figure~\ref{fig:bin_test_cmd}.~As in section  \ref{sec:mcmcresults}, we run $\mathtt{singlePopMcmc}$ with 10,000 iterations and keep the binary parameter turned off, in order to measure relative differences in the obtained parameter posterior distributions.~As it can be seen from the box plot distributions in Figure~\ref{fig:bin_test_posterior}, all three mock data sets return very similar parameter distributions.~Assuming a true binary fraction of $f_b$=0.018 \citep{milone12b} for 47 Tuc  we can thus conclude that neglecting the binaries from the analysis in section \ref{sec:mcmcresults} should not have any measurable effect on our results, which also suggests that (i) the uncertainties are dominated by random and systematic errors which arise from the photometry and the DSED models and that (ii) BASE-9 is likely classifying most binaries as field stars during the modeling, therefore decreasing their effect on the results.

\begin{figure*}[t!]
\centering
\includegraphics[width=0.7\textwidth]{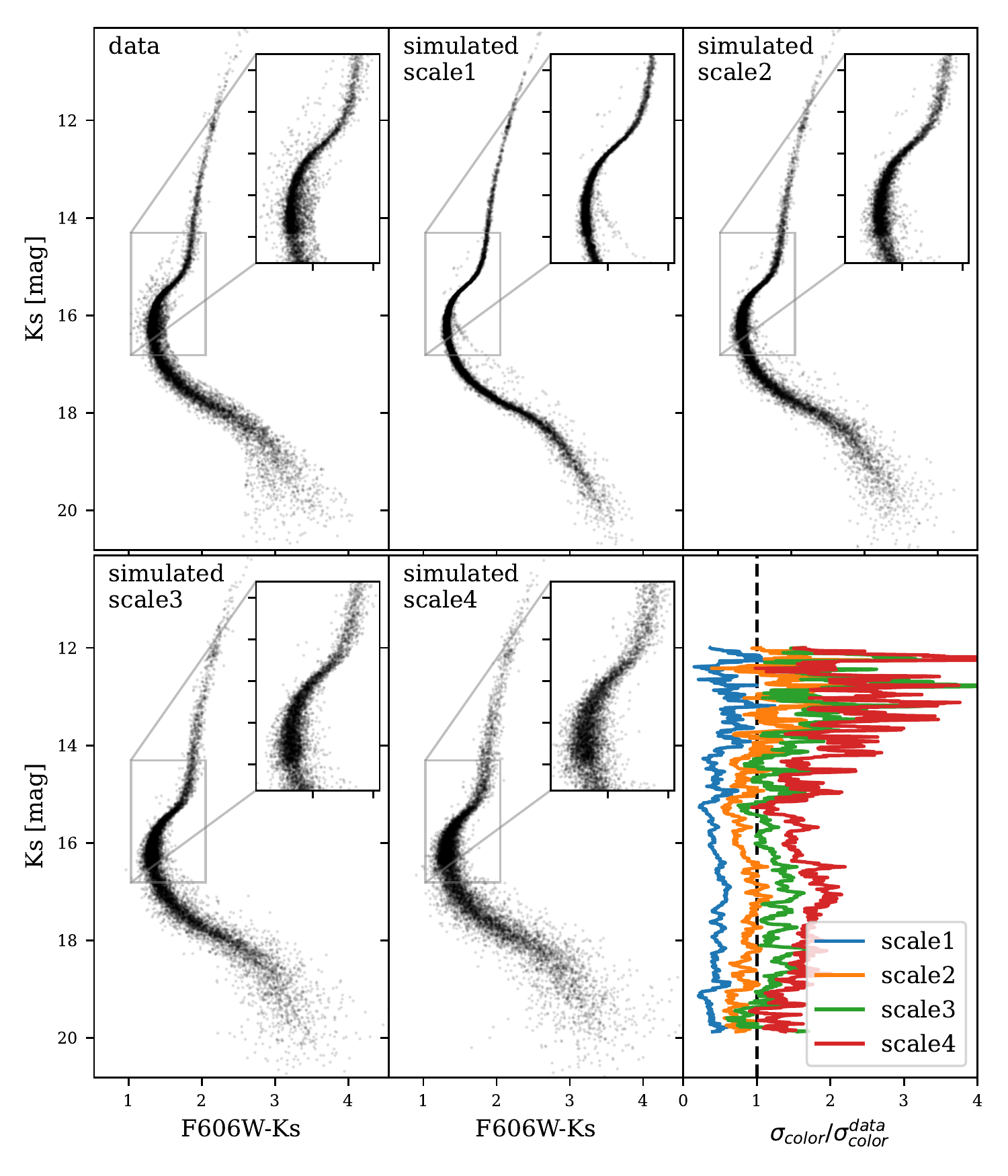}
\caption{CMD of the real photometry used in BASE-9 (top left) compared to the CMDs of mock catalogs simulated with BASE-9 for increasing noise scales (see detail in text).~The lower right panel shows the value of a running standard deviation of the color index values.~The curves are normalized to the curve obtained from the real data, hence providing a comparison of the relative broadening in the CMD as a function of  K$_s$-filter magnitude.  \label{fig:noise_test_cmd}}
\end{figure*}

\begin{figure}[t!]
\includegraphics[width=0.47\textwidth]{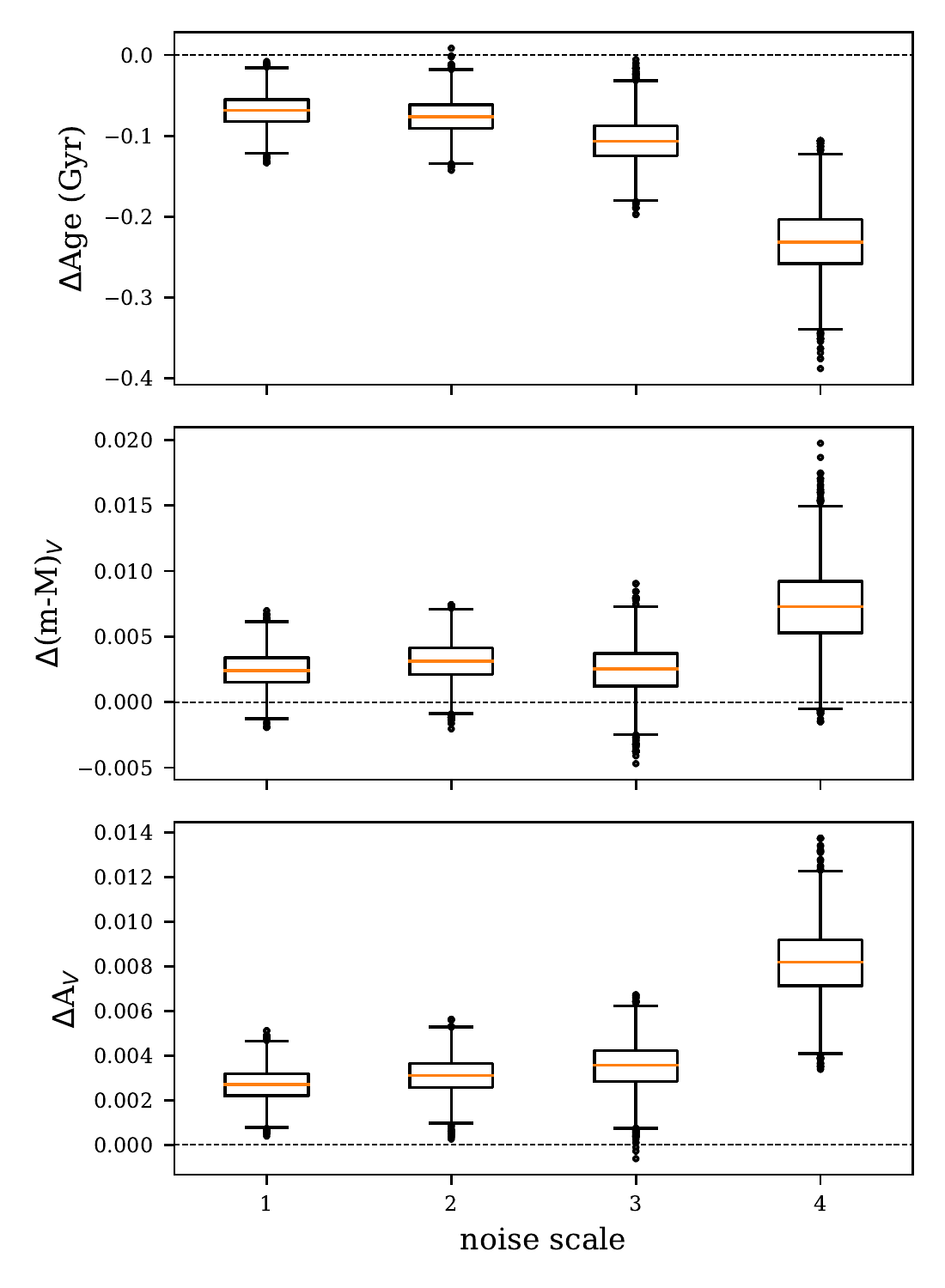}
\caption{Box plots of the derived posterior distributions for simulated catalogs assuming different noise scales as defined in Section~\ref{sec:noise_test}.~The y-axis gives the offset defined as $\Delta\theta = \theta-\theta_{true}$.~The horizontal dashed line is then perfect agreement between the true parameter and the posterior. \label{fig:noise_test_posterior}}
\end{figure}

\subsection{Effect of Noise in the Data}\label{sec:noise_test}

The next test we perform is similar to the previous one, but this time we wish to reproduce our observed data and so we have fixed the binary fraction to $f_b$ = 0.02 \citep{milone12b} and added noise to the simulated catalogs at four increasing scales.~For each each noise scale, we add noise to the simulated magnitudes by drawing from a normal distribution with $\sigma$ scaled to 1, 2, 3 and 4 times the photometric error that we get in the real data sample.~The final simulated catalogs and the real data sample are shown in Figure~\ref{fig:noise_test_cmd} for comparative purposes.~The lower right panel shows the value of a running standard deviation of the color index values.~The curves are normalized to the curve obtained from the real data, hence providing a comparison of the relative broadening in the CMD as a function of magnitude.~Note that for the binary fraction test we used a ``scale 1" model, which Figure~\ref{fig:noise_test_cmd} shows that it gives a tighter CMD than the observed data, as expected since photometric errors are not the only source of broadening in the CMD.~From the lower right panel, we find that the ``scale 2" model is able to match closely the observed width in the real data CMD, which is also confirmed visually in the CMD, especially in the zoomed inset panels.~Indeed, the SGB and MSTO sharpness looks to be fairly reproduced by the ``scale 2" model.~This is relevant for our test since these regions contain much of the diagnostic power when performing an isochrone fit.~The MCMC results for each model are shown in Figure~\ref{fig:noise_test_posterior}.~As expected, BASE-9 performs better at recovering the real cluster parameters for the ``scale 1'' and ``scale 2'' models, while for the ``scale 3'' and ``scale 4'' models the posterior distributions generally become wider and further from the true parameters.~From the plot we can infer that (i) the chain of posterior samples will not always contain the true parameter value and (ii) such systematic offset can be roughly quantified.~We can then imagine that if the ``scale 2'' model were the real data, our nominal parameter estimate (the median of the posteriors) would have an effective offset of $\Delta \theta$, as measured in the Figure.~Our simple approach is then to use the $\Delta \theta$ values from the ``scale 2" model and assume such offset as a systematic uncertainty of the results in section \ref{sec:mcmcresults}, and for simplicity we adopt them as symmetrical uncertainties.~We acknowledge that this approach is very simple, and lacks statistical robustness.~An ideal estimation of the systematics would be to repeat this test a large number of times, but the required computation time goes beyond the scope of this work.~Nevertheless, the measured offsets can still be used as an indicator for the accuracy of our MCMC sampling.

In addition, we now consider the photometric uncertainties in the determination of the zero-point for the K$_s$-band magnitude data.~From Figure~\ref{fig:zp1} we simply add both errors from the calibration and adopt $\sigma_{ZP}=$ 0.022 mag as the total uncertainty in the zero-point.~We then test for the extreme cases and create two alternative photometric catalogs by adding and subtracting $\sigma_{ZP}$ from the K$_s$ photometry in the real data catalog and calculate new posterior distributions with BASE-9.~As expected, the results show that the posteriors for age are not affected (since the structure of the CMD is unaffected), while the distance modulus and absorption do show a difference of $\pm$0.003 mag and $\pm$0.028 mag, respectively, in the posterior distribution median relative to the real data results.~We thus consider these as additional systematic uncertainties and add them to the errors derived from the noise test described above.~The final estimates with their uncertainties are listed in Table \ref{tab:final_results}.

\begin{table}
	\centering
	\caption{Calculated parameter estimates for 47 Tuc based on MCMC sampling and systematic errors.}
	\label{tab:final_results}
	\begin{tabular}{lc|r} 
		\hline
		Parameter & value & unit  \\
		\hline
		Age   & 12.42$^{+0.05}_{-0.05}$ $\pm$ 0.08 & Gyr\\
		$A_V$   & 0.090$^{+0.002}_{-0.002}$ $\pm$ 0.028 & mag\\		
		$(m-M)_V$ & 13.340$^{+0.003}_{-0.003}$ $\pm$ 0.004  & mag \\
		\hline				
	\end{tabular}
\end{table}

\section{Discussion}\label{sec:discussion}

\begin{table*}
	\centering
	\caption{Estimates of age and distance for 47 Tuc found in the literature.}
	\label{tab:age_dist_table}
	\begin{tabular}{llll} 
		\hline
		\hline
		Method & (m$-$M)$_{0}$ [mag]$^a$ & Age [Gyr] & Reference \\
		\hline
		\multirow{7}{*}{Isochrone fit to CMD} & 13.22 & 12.0 & \citet{fu18} \\
		   & 13.21$\pm$0.07 & 12.25$\,-\,$12.75 & \citet{brogaard17} \\
		   & 13.47$\pm$0.2   &  11.6$\pm$2.0  & \citet{omalley17} \\
		   & 13.30                  &    $-$                  &  \citet{niederhofer18} \\
		   & 13.273$\pm$0.004 & 13.49$^{+0.006}_{-0.022}$ & \citet{wagner17} \\
		   &  13.31$^{+0.04}_{-0.05}$  & 11.6$\pm$0.7  & \citet{correnti16}  \\ 	
		   &  13.40$\pm$0.1  & 12.25$\,-\,$12.75  & \citet{li14} \\
		   &  13.31    &  12.00                   & \citet{vandenberg14}   \\
		   & 13.26$^b$ & 12.75$\pm$0.5 & \citet{dotter10} \\
		\hline
		\multirow{2}{*}{ZAHB models + Isochrone fit} & 13.19  &  13.0  & \citet{denissenkov17} \\						
		   & 13.25 & 11.75$\pm$0.25 & \citet{vandenberg13} \\
		\hline
		\multirow{4}{*}{Model fit of detached EcB components} &  13.29$\pm$0.01  &  12.0$\pm$0.5  &  \citet{thompson20} \\ 
		 & 13.21$\pm$0.07 &  11.8$^{+1.6}_{-1.4}$ & \citet{brogaard17} \\
		 &  13.25   & 11.0$\pm$0.5  & \citet{vandenberg13}  \\ 
		 &  13.23$\pm$0.08  &  11.25$\pm$0.88 &  \citet{thompson10} \\  
		 \hline		
		\multirow{3}{*}{WD cooling sequence} &  13.20$^c$ & $\sim$12.0 & \citet{garciaberro14} \\
		  &  13.32$\pm$0.09 & 9.9$\pm$0.7 & \citet{hansen13} \\
		  & 13.36$\pm$0.06 & $-$ & \citet{woodley12} \\
		  \hline		

		\multirow{2}{*}{Dynamical modeling from HST proper motions} & 13.18$\pm$0.24 & $-$ & \citet{heyl17} \\
		   &  13.09$\pm$0.04  &  $-$  &  \citet{watkins15} \\ 		   
		\hline

                  \multirow{2}{*}{Gaia DR2 parallax}  &  13.25$\pm$0.24  & $-$  & \citet{shao19} \\
		   &  13.24$\pm$0.06  & $-$  & \citet{chen18} \\
		   &   13.239$\pm$0.002  &  $-$ & \citet{gaia18} \\
		\hline
		
		\multicolumn{4}{l}{$^a$ (m$-$M)$_V$ values were dereddened assuming $R_V=A_V/E(B-V)=3.1$ and their reported value for $E(B-V)$.}\\
		\multicolumn{4}{l}{$^b$ Transformed from (m$-$M)$_{F814W}$ using the extinction ratios from Table A1 in \citet{casagrande14}.} \\
		\multicolumn{4}{l}{$^c$ Not originally reported in the cited paper, but reported by \citet{brogaard17} through private communication. }\\
		\multicolumn{4}{l}{Note - If random and systematic errors are listed, we have assumed they are uncorrelated and thus combined in quadrature.}
	\end{tabular}
\end{table*}

\begin{figure}[ht!]
\includegraphics[width=0.47\textwidth]{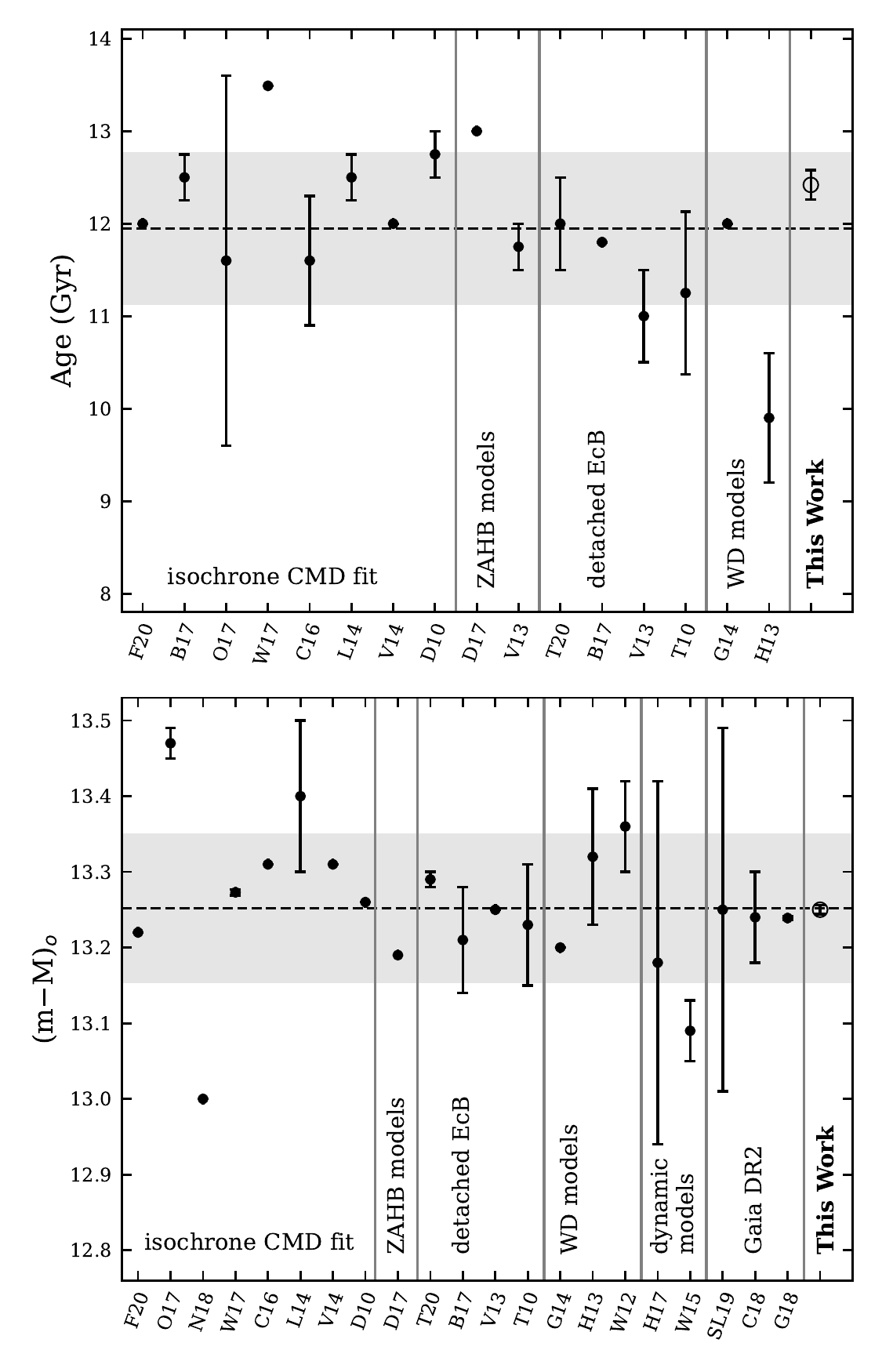}
\caption{Age (top) and distance (bottom) estimates of 47 Tuc found in the literature.~The methods used are labeled in the figure.~We show also the mean value and the 1$\sigma$ range (dashed line and gray region respectively).~The right most symbol gives our estimated values.~Note that in our case (m$-$M)$_0$ is of course calculated from (m$-$M)$_V$ and A$_V$ as given in Table~\ref{tab:final_results}.\\
~References: B17:~\citet{brogaard17}, C16:~\citet{correnti16}, C18:~\citet{chen18}, D10:~\citet{dotter10}, D17:~\citet{denissenkov17}, F20:~\citet{fu18}, G14:~\citet{garciaberro14}, G18:~\citet{gaia18}, H13:~\citet{hansen13}, H17:~\citet{heyl17}, L14:~\citet{li14}, N18:~\citet{niederhofer18}, O17:~\citet{omalley17}, SL19:~\citet{shao19}, T20:~\citet{thompson20}, T10:~\citet{thompson10}, V14:~\citet{vandenberg14}, V13:~\citet{vandenberg13},  W17:~\citet{wagner17}, W12:~\citet{woodley12}, W15:~\citet{watkins15}   \label{fig:age_dist_lit}}
\end{figure}

\subsection{The distance and absolute age of NGC 104}\label{sec:distance_and_age}
47 Tuc is one of the most studied Galactic GCs due to its relative proximity and, hence, apparent brightness.~It has been widely used as a calibrator for models of metal-rich stellar populations which are used to study observed populations in nearby galaxies, and GCs in the local universe.~There have been numerous studies published in the last decade which have measured its heliocentric distance and absolute age, and here we try to provide, to the best of our knowledge, a detailed compilation of these studies from the last decade.~The summarized literature values are presented in Table~\ref{tab:age_dist_table}.~Note that some studies have been listed twice, which means that they have presented separate results using independent methods.~For the interested reader, we refer to \citet[see their Table~1]{woodley12} for a compilation of heliocentric distance estimates of 47 Tuc published until 2010. 

One common approach to derive cluster parameters is to perform isochrone fits to the main sequence and RGB in the CMD.~Using the high-quality optical HST photometry from \citet{sarajedini07}, distance and age was obtained by \citet{dotter10} and more recently by \citet{wagner17}, \citet{omalley17}, \citet{brogaard17} and \citet{vandenberg14}.~Their results show a range in ages between $\sim$11.6$-$12.75 Gyr, except for \citet{wagner17} who finds a considerably older age of $\sim$13.5 Gyr.~However, such old age can be explained given that their study used [$\alpha$/Fe]=0.0 models and simultaneously fixed the metallicity to [Fe/H]=$-$0.72, which likely results in models with a lower $\alpha$-element abundance than what is appropriate for 47 Tuc, considering that spectroscopic studies have found its $\alpha$-enhancement to be closer to [$\alpha$/Fe]~$\sim$~0.3$-$0.4 \citep{carretta10,carretta09b,koch08,mcwilliam08,alvesbrito05}.~The driving factor is the lower [O/Fe] abundance, which results in a brighter and hotter MS turn-off point, due to its important role in the CNO-cycle.~The isochrone fitting will then have to compensate by using an older age for its best-fit model.~Also noteworthy is the work by \citet{fu18} who used the very deep (down to $\sim$30th magnitude) HST optical photometry from \citet{kalirai12}.~While also using $\alpha$-enriched models, and appropriate metallicity and helium fraction for first and second generation stars, they derive from isochrone fitting an age of 12.0 Gyr and true distance modulus (m$-$M)$_0$=13.22 mag.

Also using isochrone fitting, other recent studies have instead used near-IR photometry CMDs to derive cluster age and distance.~\citet{niederhofer18} and \citet{li14} both used the same near-IR wide-field imaging data from the VMC survey \citep{cioni11}.~They find true distance modulus values of 13.30 and 13.40, respectively.~While \citet{niederhofer18} assumed a fixed single age of 11.8 Gyr for the isochrone fit, \citet{li14} derived the age by fitting isochrones for a metal-poor, helium-poor population, and separately for a metal-rich, helium-rich population.~They find an older age of 12.75 Gyr for the first, and 12.25 Gyr for the metal-rich and younger population.~Another interesting work is that of \citet{correnti16}, who used very deep near-IR HST photometry to fit the upper main-sequence as well as the lower main-sequence, including the LMSD at high photometric accuracy.~They used a grid of models and only fixed one parameter, [$\alpha$/Fe]=0.4, and adjusted the helium appropriately by following a helium enrichment law of $\Delta Y$/$\Delta Z$=1.5, in correspondence with the different metallicity values in the isochrone grid.~The best fit age was found to be 11.6 Gyr and true distance modulus (m$-$M)$_0$=13.31 mag.~The above set of studies therefore suggest an absolute age somewhere between 11.6 and 12.75 Gyr, for the main-sequence isochrone fit method.

The detached eclipsing binary (EcB) V69 in 47 Tuc \citep{weldrake04} gives another opportunity to measure precise cluster parameters.~Once the stellar parameters of the V69 components are derived with spectroscopic radial velocities and photometric light curves, the luminosity can be determined and hence the absolute magnitude and distance modulus.~Then, mass-radius and mass-luminosity relations are used to derive the absolute age of the components.~This was first done by \citet{thompson10}, and later by \citet{vandenberg13}, \citet{brogaard17} and \citet{thompson20}, who also include a second additional detached EcB, E32.~Their results, each depending on particular chemical composition assumptions, are mildly consistent with each other and around 11.0$-$12.0 Gyr for the age, and around 13.21$-$13.29 for the true distance modulus (m$-$M)$_0$.~The large age uncertainty (11.8$^{+1.6}_{-1.4}$ Gyr) shown in Table~\ref{tab:age_dist_table} for \citet{brogaard17} is their conservative 3$\sigma$ limit which covers all acceptable solutions that satisfy the mass-radius, mass-luminosity and CMD conditions.~Indeed, as \citet{thompson20} has recently pointed out with respect to V69, there is a broad range of possible ages allowed if the CMD, mass-radius and mass-luminosity relations must be all satisfied, and thus additional spectra with high S/N are needed to obtain direct determination of temperature and metallicity of the detached EcB components. 

Another recent pair of studies have used zero-age horizontal-branch (ZAHB) models combined with MSTO isochrone fitting to derive distance and age to 47 Tuc.~\citet{vandenberg13} obtained an age of 11.75 Gyr and (m$-$M)$_0$=13.25 mag, although their distance modulus was independently derived from the stellar parameters of the detached EcB V69 by using the data from \citet{thompson10}.~More recently, \citet{denissenkov17} used ZAHB models to fit the observed HB in 47 Tuc assuming three populations with different helium composition, and obtained a true distance modulus (m$-$M)$_0$=13.19 mag.~This value seems small compared with most other studies in Table~\ref{tab:age_dist_table}, and an underestimation would be in line with their relatively high estimate for the age, at 13.0 Gyr.~Another factor that could bias towards older age is the fact that for their ZAHB models to match the observations, they had to adopt a relatively small extinction, E($B-V$)=0.026, which is smaller than most values found in the literature, e.g. around 0.032 \citep{schlegel98} and 0.04 \citep{harris10}.

The WD population of 47 Tuc was observed with HST imaging by \citet{woodley12} to build WD SEDs and compare them to theoretical WD atmosphere models, yielding a true distance modulus (m$-$M)$_0$=13.36 mag.~The same dataset was then used by \citet{hansen13} to obtain a similar distance and, by including WD cooling models into their analysis, an age of 9.9 Gyr, which is the youngest estimate in Table~\ref{tab:age_dist_table}.~A similar analysis with the same HST data was done later by \citet{garciaberro14}, who intriguingly obtained a smaller distance and consequentially a much older age than the previous authors, at around 12 Gyr.

Other methods can only provide a distance estimate.~The studies by \citet{watkins15} and \citet{heyl17} are based on combining HST proper motions and radial velocities to generate kinematic models of 47 Tuc and derive its distance.~We note that both provide the two shortest distance modulus estimates available in Table~\ref{tab:age_dist_table}: 13.09 and 13.18, respectively.~Interestingly, an earlier kinematic distance estimate from \citet{mclaughlin06} is even smaller at (m$-$M)$_0$=13.02.~This suggests that there is something inherent to the kinematic modeling that might result in shorter distance estimates than those derived by e.g. main-sequence isochrone fitting,  detached EcBs, WDs, etc, yet a more detailed explanation for this effect goes beyond the scope of this paper. 

More recently, as Gaia DR2 catalogs \citep{gaia18} have become available, parallax-based distances to 47 Tuc have been published first by \citet{gaia18} and \citet{chen18}, both giving remarkably consistent estimates at (m$-$M)$_0$=13.24 mag.~A later work by \citet{shao19} used a mixture model to fit the cluster mean parallax, as well as an additional field star component, thus avoiding the problem of membership determination and maximizing the Gaia sample size.~Their estimate is also consistent at (m$-$M)$_0$=13.25 mag.~We note that this consistency nicely supports the detached EcB-based distances in Table~\ref{tab:age_dist_table}, which have an arithmetic mean $\langle$(m$-$M)$_0\rangle=13.245$ mag with standard deviation $\sigma$ = 0.03 mag.  

We show in Figure~\ref{fig:age_dist_lit} the distribution of values for age and distance and its arithmetic mean value (black dashed line) and standard deviation (gray shaded region), as well as the methods used in each measurement.~The estimates from this work are shown on the far right with an open circle symbol (and are not used for the mean).~For the age, the figure gives us a mean value $\approx$12 Gyr which is slightly younger than our estimated age of 12.42$^{+0.05}_{-0.05}$ $\pm$ 0.08 Gyr.

~For the true distance modulus, we of course calculate it as (m$-$M)$_0$ = (m$-$M)$_V$ $-$ A$_V$, from the values in Table~\ref{tab:final_results}.~Our estimate is then\footnote{We note that for calculating the errors it needs to be accounted that distance modulus and absorption are correlated, and so their covariance was calculated from the MCMC sampling: cov(m$-$M$_V$,A$_V$)$\sim$10$^{-6}$.}, (m$-$M)$_0$=13.250$^{+0.003}_{-0.003}$ $\pm$ 0.028, which is remarkably close to the mean value of the figure: $\langle$(m$-$M)$_0\rangle$=13.252, and also in good agreement (within $<$0.01 mag) to the mean from Gaia DR2 measurements: $\langle$(m$-$M)$_0\rangle$=13.243, and the mean from detached EcB measurements: $\langle$(m$-$M)$_0\rangle$=13.245.~These two methods appear to provide a higher internal consistency, while the values derived from isochrone fit, WD models and dynamical modeling show larger scatter, with each other and internally.~We hence suggest that the true distance to 47 Tuc is likely well constrained by our result together with the set of estimates from Gaia DR2 and detached EcB.~Our estimate can be converted to heliocentric distance as $d = 4.47^{+0.01}_{-0.01}\pm0.08$ kpc.   

\begin{table*}[t!]
	\centering
	\caption{Estimates for metallicity and $\alpha$-enhancement of 47 Tuc from high resolution spectroscopy.}
	\label{tab:metal_literature}
	\begin{tabular}{lllll} 
		\hline
		\hline
		Reference & [Fe\,I/H] & [$\alpha$/Fe]$^{a}$ & [M/H]$^{b}$ & Data \\
		\hline  
		\citet{colucci17} & $-$0.65 $\pm$ 0.05  & 0.29 $\pm$ 0.09 & $-$0.44 & Du Pont echelle integrated-light spectra \\
		\citet{cordero14} &  $-$0.79 $\pm$ 0.09  & 0.29 $\pm$ 0.06 &  $-$0.58  &  VLT-FLAMES/Blanco-Hydra RGB spectra\\
		\citet{thygesen14} &  $-$0.78 $\pm$ 0.07  & 0.34 $\pm$ 0.03  & $-$0.54  & VLT-UVES RGB spectra \\
		\citet{gratton13} &  $-$0.76 $\pm$ 0.01  & 0.29 $\pm$ 0.10 & $-$0.55 & VLT-FLAMES HB spectra \\ 
		\citet{carretta09c,carretta10} &  $-$0.76 $\pm$ 0.02  & 0.42  &  $-$0.45 & VLT-UVES RGB spectra\\
		\citet{koch08} & $-$0.76 $\pm$ 0.01 $\pm$ 0.04  & 0.39 $\pm$ 0.04 &  $-$0.47 & Magellan-MIKE RGB spectra \\
		\citet{mcwilliam08} & $-$0.75 $\pm$ 0.026 $\pm$ 0.045  & 0.36 $\pm$ 0.12 & $-$0.49  &  Du Pont echelle integrated-light spectra \\
		\citet{alvesbrito05} &  $-$0.66 $\pm$ 0.07  & 0.19 $\pm$ 0.11  &  $-$0.53 &  VLT-UVES spectra of RGB \\
		
				     & \multicolumn{3}{c}{Mean $\pm$ $\sigma$ }  &  \\
		 \hline
		                            &  $-$0.74 $\pm$ 0.05   &  0.32 $\pm$ 0.07   &  \multicolumn{2}{l}{$-$0.50 $\pm$ 0.05}   \\
		 \hline		
		\multicolumn{5}{l}{Notes - ($a$) Mean([X/Fe]) $\pm$ $\sigma$ of elements Mg, Ca, Si and Ti when [$\alpha$/Fe] not explicitly given in the paper.}\\
		\multicolumn{5}{l}{ ($b$) Calculated using Equation~\ref{eq:metal}.
		}		
	\end{tabular}
\end{table*}

\begin{figure}[t!]
\includegraphics[width=0.45\textwidth]{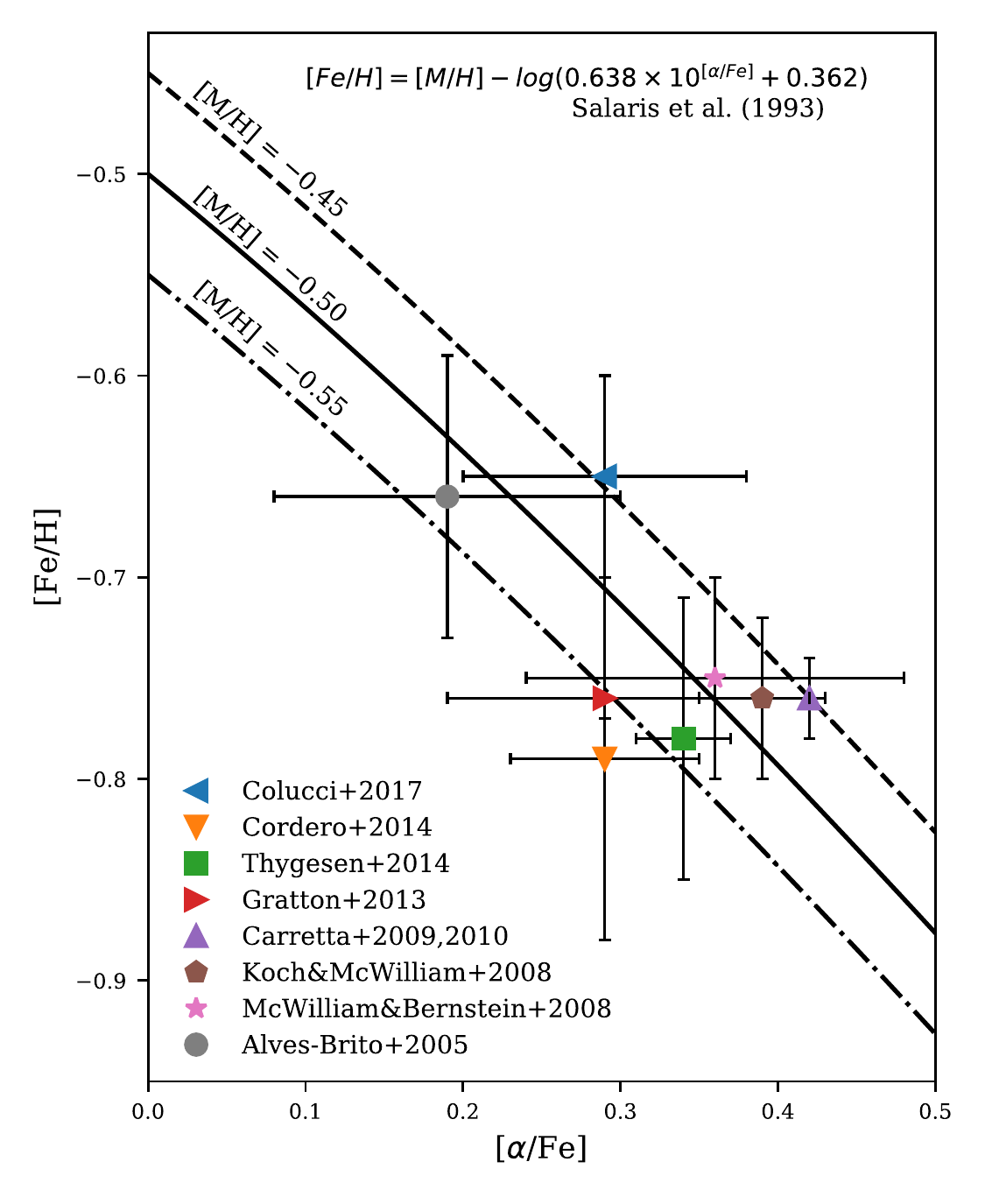}
\caption{Estimates of iron-abundance and alpha-enhancement of 47 Tuc obtained from high resolution spectroscopy in previous literature.~We show the relation between [Fe/H] and [$\alpha$/Fe] for a fixed value of global metallicity [M/H].~The solid black line corresponds to our fixed prior value on global metallicity [M/H] = $-$0.5. \label{fig:alpha_Fe_relation}}
\end{figure}

\begin{figure*}[ht!]
\centering
\includegraphics[]{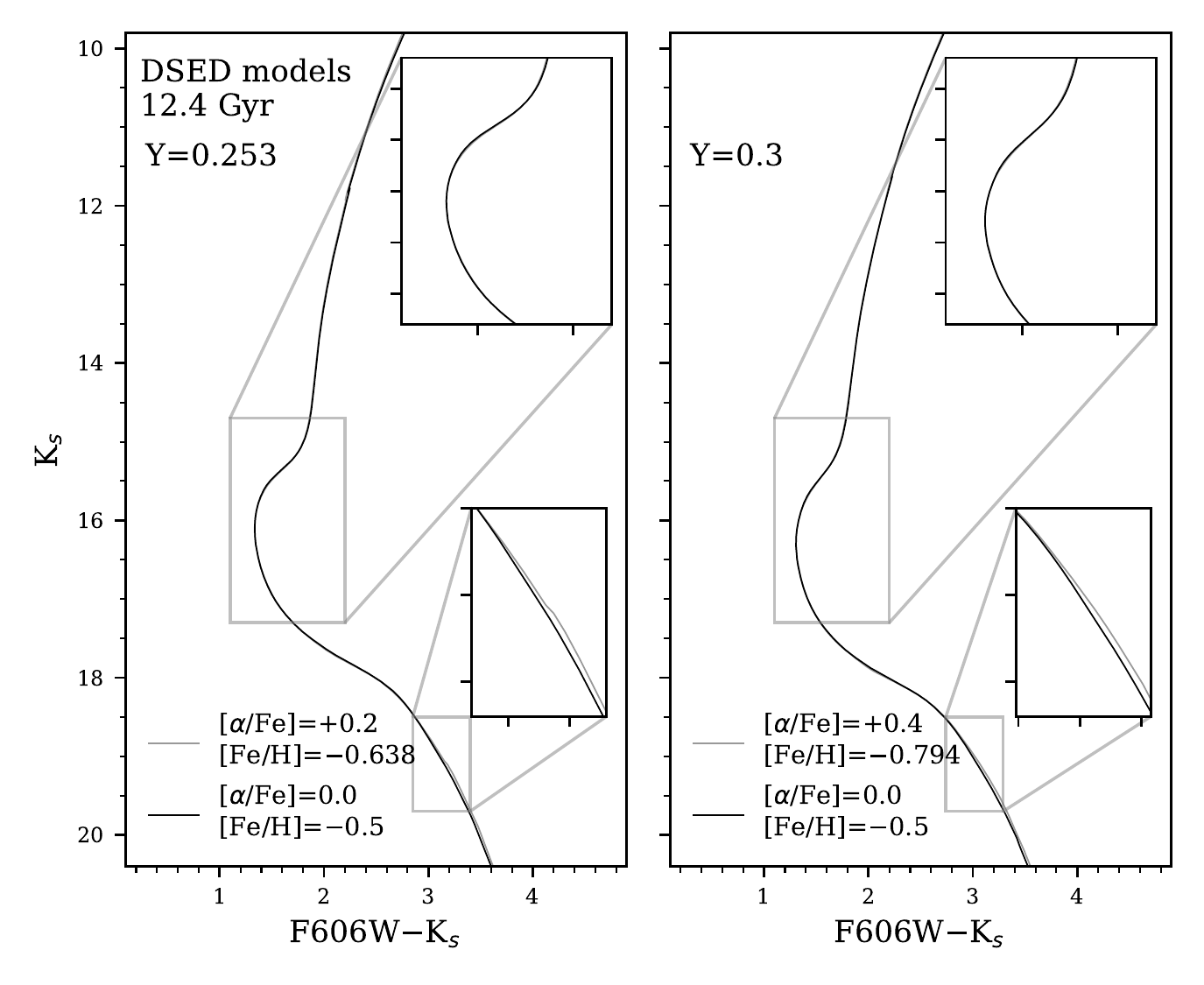}
\caption{Isochrone models from the DSED library created with equal global metallicity, age, distance and absorption parameters as our results in Table~\ref{tab:final_results}.~The color and magnitude axis range are equal to Figure~\ref{fig:vk_isochrone} for easier comparison.~In each panel, we plot two isochrones: without alpha-enhancement (black line) and with alpha-enhancement (gray line).~The iron abundance is chosen so to keep [M/H] constant at [M/H]=$-$0.50.~The two isochrones are mostly indistinguishable on the plot.~We make this test for different helium mass fraction values: $Y$= 0.253 (left panel) and $Y$= 0.3 (right panel).  \label{fig:alpha_enhanced_CMD}}
\end{figure*}

\subsection{The Metal Mixture Content of 47 Tuc}\label{sec:metal_content}

As described earlier, we argue that global metallicity [M/H] is the preferred parameter to describe metal content for when no $\alpha$-enhanced models are available, such as in our case.~In order to aggregate the estimates from the literature, we need to derive [M/H] values from published spectroscopic values of [$\alpha$/Fe] and [Fe/H].~We list in Table~\ref{tab:metal_literature} a compilation of measurements from high-resolution spectroscopy.~Note that the $\alpha$-enhancement is sometimes not explicitly given in each paper, and for these cases we have calculated its value from the average element-to-iron abundance of $\alpha$-elements (Mg, Ca, Si, Ti)\footnote{Oxygen is ignored from the calculation due to the large dispersion that is present in RGB stars as a result of the Na-O anticorrelation. Note also that, when possible, we use the Ti II abundance alone and ignore neutral Ti because of discrepancies due to NLTE effects \citep{bergemann11,thygesen14}.}.~The table shows good agreement within most studies, placing their likely values in between $-$0.79 $\lesssim$ [Fe/H] $\lesssim$ $-$0.75 and 0.3 $\lesssim$ [$\alpha$/Fe] $\lesssim$ 0.4.~For each case, the corresponding derived global metallicity is shown in the fourth column.~The average value is $\langle$[M/H]$\rangle$ = $-$0.50 with standard deviation $\sigma$ = 0.05, and this is used as the fixed prior value in Section \ref{sec:iso_fit}.~In Figure~\ref{fig:alpha_Fe_relation} we show how different pair values of [Fe/H] and [$\alpha$/Fe] are linked to a specific global metallicity.~The plot visually illustrates that although metal abundance estimates can vary, a global metallicity [M/H]=$-0.5$ can explain the bulk of the literature within the uncertainties.~We thus highlight that studies relying on stellar evolutionary tracks should not neglect using suitable $\alpha$-enhancement in their models and take special caution when assuming [Fe/H]$\approx$[M/H].

This brings up the question of whether we need to correct for the fact that our DSED models are solar-scaled, i.e. [$\alpha$/Fe]=0.~Fortunately, the isochrone model should not change significantly in color-magnitude as long as the stellar radiative opacity is similar, within the specific wavelength bands, to an $\alpha$-enhanced model with equal global metallicity.~That is, the isochrones we use can be equivalent to the $\alpha$-enhanced isochrones for a fixed global metallicity.~This is illustrated in Figure~\ref{fig:alpha_enhanced_CMD} where for [M/H]=$-$0.50 we plot two DSED isochrones\footnote{$\alpha$-enhanced isochrones must be downloaded directly from the DSED Web Tool isochrone generator at http://stellar.dartmouth.edu/models/isolf\_new.html} in each panel: one for [$\alpha$/Fe]=0.0 and one for the $\alpha$-enhanced case.~The [Fe/H] values are chosen such that the global metallicity remains constant following Eq.~\ref{eq:metal}.~The left and right panels correspond to different values of helium mass fraction, as labeled on the top-left of each panel.~The isochrones have been shifted to the best-fit absorption and distance values found for 47 Tuc and the axes range is set equal to Figure~\ref{fig:vk_isochrone} for easier comparison.~The figure clearly shows that the two isochrones are mostly undistinguishable by eye in the (F606W$-$K$_s$) vs K$_s$ parameter space, at least within the color-magnitude range that was used in our analysis.~It follows then that, for DSED isochrones in the (F606W$-$K$_s$) vs K$_s$ CMD, we can find an $\alpha$-enhanced isochrone model \textit{with identical age, [M/H], Y, distance and absorption parameters} that can successfully fit our photometric data, at least to the precision shown by the figure.~This provides confidence in that all the cluster parameters derived in our analysis are robust.~We emphasize that using solar-scaled isochrones might not be adequate when comparing to the photometric bands in which the opacity can be more significantly affected by the $\alpha$-elements, e.g. in the I-band where cool stars exhibit the strong CaII triplet at $\sim$8500 \AA.~Detailed discussions on the effect of $\alpha$-enhanced mixture in stellar model computations can be found in e.g. \cite{pietrinferni06}, \cite{pietrinferni21}, \cite{cassisi04}.

\subsection{The Helium Content of 47 Tuc}\label{sec:helium_content}
The literature about He abundance in 47 Tuc is mostly in agreement on the fact that the multiple populations phenomenon in 47 Tuc manifests itself through a star-to-star He spread that is radially dependent.~The idea of He spread in 47 Tuc was first proposed by \citet{briley97} from his study of the CN-band strength bimodality in RGB and HB stars.~A later study by \citet{anderson09} concluded that the MS broadening observed in HST data could be explained solely by a He spread if $\Delta Y\sim0.026$.~Similarly, \citet{milone12} used HST data to separate multiple sequences throughout the CMD, from MS to the HB, and linked the two major sequences to CN-strong, Na-rich, O-poor, He-rich ($\Delta Y\sim0.02$) and CN-weak, Na-poor, O-rich, primordial He populations, which are associated with second generation (SG) and first generation (FG) stars, respectively.~In addition, \citet{dicriscienzo10} found that the morphology of the HB requires $\Delta Y=0.02$ to be explained by the models, a result confirmed by \citet{nataf11} who uses a centrally concentrated He-enriched ($\Delta Y=0.03$) SG population to explain the morphology of the RGB bump and the HB, and also confirmed by \citet{gratton13} who concludes that a spread $\Delta Y\sim0.02-0.03$ is inferred from the HB shape.~Another study by \citet{li14} used wide-field deep photometry to study the radial gradients in FG and SG stars.~They find a radial gradient in He which changes from $Y$=0.28 in the center to $Y$=0.25 in the outskirts.~A detailed synthetic HB modelling analysis by \cite{salaris16} also confirms $\Delta Y=0.03$ is needed to properly reproduce the HB morphology.~More recently, \citet{denissenkov17} used state-of-the-art HB models to quantitatively constrain the He spread by populations and found again $\Delta Y=0.03$, with 21\%, 37\% and 42\% of stars having $Y$=0.257, $Y$=0.27 and $Y$=0.287, respectively. 

We therefore emphasize that the adopted fixed prior for helium, $Y$=0.28, is not to be interpreted as representative of the entire cluster population.~Rather, depending on the studied sample, the choice for Y should be considered as an average of the He abundances of the observed population.~The fraction of He-rich SG stars in 47 Tuc is about 70\% \citep{carretta09a,carretta10,gratton13} and increases to about 80\% in the central regions \citep{li14}.~Given that our data encompasses only the inner $\sim$85\arcsec (i.e. out to $\sim$0.2\,$r_h$) we expect that the large majority of the stars observed are He-rich stars, and hence our choice for the Y prior.

\section{Summary and Conclusions}\label{sec:conclusions}

This is the second published work of our ongoing G4CS project to obtain homogenous and deep near-IR photometry of Galactic GCs.~These data will allow us to probe deep into the low MS of GCs and enhance our ability to obtain precise and accurate absolute age estimates, as well as a full characterization of the structural parameters.~The degeneracies that exist between certain cluster parameters is a fundamental limitation of these kinds of studies, and thus high-quality data that can constrain specific theoretical models down to their inherent precision is of absolute importance.

For this, the photometric catalogs of HST GC surveys are some of the best data sets available due to their impressive spatial-resolution and photometric accuracy.~Thus, the goal of G4CS is also to provide the near-IR data catalog counterparts at similar instrumental precision, while additionally creating a new window to improve on the stellar evolutionary models at near-IR wavelengths, and their internal consistency for various near-UV+optical+near-IR combinations.~For this work, we have combined HST optical F606W-band and GSAOI K$_s$-band magnitudes to study the CMD of 47 Tuc and characterize its structural parameters.

    We presented a proof of concept regarding the use of Bayesian statistics to precisely fit the photometric data of an optical-nearIR CMD, and showed that the obtained isochrone model, with the exception of HB stars which we did not include in the analysis, can successfully reproduce the sharp SGB and MSTO structures, the upper and lower MS and the RGB.~Our Bayesian MCMC analysis allowed us to obtain probability distributions and highly precise estimates of our free model parameters.~In future work, we hope to transition into using pure near-IR (J, K$_s$) catalogs, where the low MSK can be used at its full potential for isochrone fitting, but we remain limited by the observational challenges of achieving a good PSF in the J-band.~We summarize this paper as follows:

\begin{enumerate}
\item A search in the literature for high-resolution spectroscopy studies reveals consistent values for the metal content of 47 Tuc.~The set of measurements give an average global metallicity of [M/H]= $-$0.5, and the individual studies are generally consistent with this average when taking uncertainties into account.~Our model prior is fixed to this value.

\item A search in the literature for estimates of Helium content shows consensus that 47 Tuc has an internal spread of He abundance up to $\Delta Y$$\sim$0.03, with a centrally concentrated Helium-rich ($Y\sim0.28$) population likely associated to SG stars.~For fitting our data of the central field of the cluster, we use a fixed model prior of $Y$=0.28. 

\item Based on the MCMC sampling of the posteriors, our best age estimate is 12.42$^{+0.05}_{-0.05}$ $\pm$ 0.08 Gyr; a value well within the range of previous estimates which are scattered around $\sim$12 Gyr. The best extinction estimate is A$_V$ = 0.090$^{+0.002}_{-0.002}$ $\pm$ 0.028 mag.

\item The mean distance modulus from a set of Gaia DR2 parallax studies $\langle$(m$-$M)$_0\rangle$=13.243 mag is tightly consistent with the mean value from detached EcBs studies $\langle$(m$-$M)$_0\rangle$=13.245 mag.~Our best estimate is also in good agreement: (m$-$M)$_0$=13.250$^{+0.003}_{-0.003}$ $\pm$ 0.028 mag, which adds more confidence to our age estimate in particular, given their strong correlation. 

\end{enumerate}

\begin{acknowledgments}

The authors thank the anonymous referee for their helpful comments that improved the quality of the manuscript.~M.S. would like to thank Ted Von Hippel for his great support on the use and configuration of the BASE-9 software.~Also would like to thank Mischa Schirmer for his help with the THELI program.~T.H.P. acknowledges support through FONDECYT Regular project 1201016 and CONICYT project Basal AFB-170002.

Based on observations obtained at the international Gemini Observatory, a program of NSF’s NOIRLab, which is managed by the Association of Universities for Research in Astronomy (AURA) under a cooperative agreement with the National Science Foundation. on behalf of the Gemini Observatory partnership: the National Science Foundation (United States), National Research Council (Canada), Agencia Nacional de Investigaci\'{o}n y Desarrollo (Chile), Ministerio de Ciencia, Tecnolog\'{i}a e Innovaci\'{o}n (Argentina), Minist\'{e}rio da Ci\^{e}ncia, Tecnologia, Inova\c{c}\~{o}es e Comunica\c{c}\~{o}es (Brazil), and Korea Astronomy and Space Science Institute (Republic of Korea). 
\end{acknowledgments}

%

\vspace{5mm}
\facilities{HST(ACS), Gemini(GSAOI),Gemini(GeMS)}






\bibliography{MyBibliography}{}
\bibliographystyle{aasjournal}



\end{document}